# Principal Decomposition of Velocity Gradient Tensor in the Cartesian Coordinates

Chaoqun Liu*, Yifei Yu, Yisheng Gao
University of Texas at Arlington, Arlington, Texas 76019, USA

cliu@uta.edu

**Abstract**

Traditional Cauchy-Stokes decomposition of velocity gradient tensor gives a symmetric and an anti-symmetric subtensors which are called the strain and vorticity tensors. There are two problems with Cauchy-Stokes decomposition. The first one is that the anti-symmetric (vorticity) tensor cannot represent the fluid rotation or vortex. The second one is that the symmetric (strain) tensor cannot distinguish the stretching (compression) and shear. The stretching and shear are dependent on the coordinate or are not Galilean invariant. Since vorticity cannot distinguish between the non-rotational shear and the rigid rotation, vorticity has been decomposed to a rigid rotation called "Liutex" and anti-symmetric shear in our previous work. A Liutex-based principal coordinate was developed and the velocity gradient tensor was decomposed in the principal coordinate as a rigid rotation (Liutex tensor), a pure shear tensor and a stretching (compression) tensor, which is called the principal decomposition. However, the principal decomposition is made at each point which has own principal coordinate different from other points. This paper derives the principal decomposition in the original xyz- coordinate system, and, therefore, provides a new tool for fluid kinematics to conduct the velocity gradient tensor decomposition to rigid rotation, pure shear, and stretching (compression) which is unique, Galilean invariant and has clear physical meanings. The new velocity gradient tensor decomposition could become a foundation for new fluid kinematics.

1. **Introduction**

Cauchy-Stokes velocity gradient tensor decomposition, i.e. $\nabla \vec{v} = A + B = \frac{1}{2}(\nabla \vec{v} + \nabla \vec{v}^T) + \frac{1}{2}(\nabla \vec{v} - \nabla \vec{v}^T)$, which is equivalent to Helmholtz velocity decomposition, has been widely accepted as the foundation of fluid dynamics for long time (Truesdell 1954 [1]). However, there are several problems with these decompositions, which cannot be neglected. Galilean invariance is a property first described by Galilei [2] in 1632, indicating whether a quantity changes under different coordinates. Any objective quantity that does not rely on the choice of coordinates must be Galilean invariant. Unfortunately, shear and stretching in Cauchy-Stokes decomposition are not Galilean invariant which means under different coordinate frames, the elements of the velocity tensor are varying with the frame rotation, showing stretch (compression) and shear are dependent on the angle of the frame rotation when $\nabla \vec{V} = \boldsymbol{Q} \nabla \vec{v} \boldsymbol{Q}^T$, where $\boldsymbol{Q}$ is an orthogonal and rotation matrix (Liu et al., 2018 [3]; Gao et al., 2018 [4] and Liu et al. 2019 [5]; Yu et al., 2020 [6]). Another problem is the anti-symmetric part of the velocity gradient tensor is not the proper quantity to represent fluid rotation as discussed by Kolar (2007) [7] and our previous work. Helmholtz's velocity decomposition divided the fluid velocity to two parts, claimed that one has potential (deformation but no rotation) and the other has rotation (vorticity but no divergence), which has equivalent tensor version called Cauchy-Stokes decomposition. These decompositions are correct in mathematics. However, these decompositions induce many questions as vorticity has no correlation with flow rotation and $\nabla \times \vec{v} \neq 0$ does not mean the flow is rotational. On the other hand, vorticity has part of anti-symmetric shear and strain has part of symmetric shear (Kolar, 2007) [7]. Helmholtz (1858) [8] first considered a vorticity tube with infinitesimal cross-section as a vortex filament, which was followed by Lamb (1932) [9] to simply call a vortex filament as a vortex in his book. Since vorticity is well-defined, vorticity dynamics has been systematically developed for the generation and evolution of vorticity and applied in the study of vortical-flow stability and vortical structures in transitional and turbulent flows (Wu et al., 2006) [10] ; Saffman, 1992 [11]; Majda et al., 2001 [12]). However, the use of vorticity will meet severe difficulties in viscous flows, especially in turbulence: (1) vorticity is unable to distinguish between a real vortical region and a shear layer region; (2) it has been noticed by several researchers that the local vorticity vector is not always aligned with the direction of vortical structures in turbulent wall-bounded flows, especially at locations close to the wall (Zhou et al., 1999 [13]; Pirozzoli et al., 2006 [14]; Gao et al., 2011 [4]) (3) the maximum vorticity does not necessarily occur in the central region of vortical structures (Robison, 1991 [15]; Wang et al., 2017 [16]). Fig.1 and 2 exhibits a direct numerical simulation result of boundary layer transition. It shows clearly that the streamlines rotates around Liutex core line rather than vorticity line.

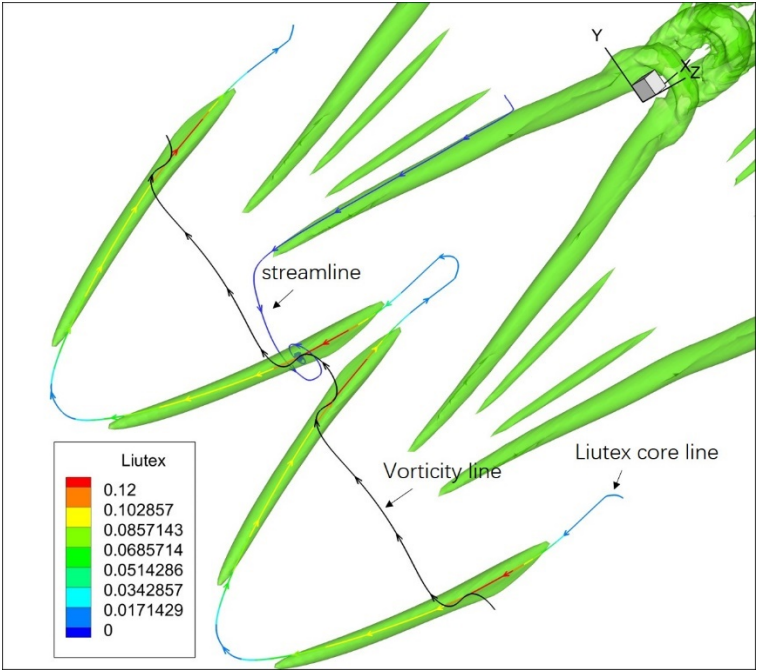

Fig. 1 Liutex core line, vorticity line and streamline with Liutex iso-surface R=0.07

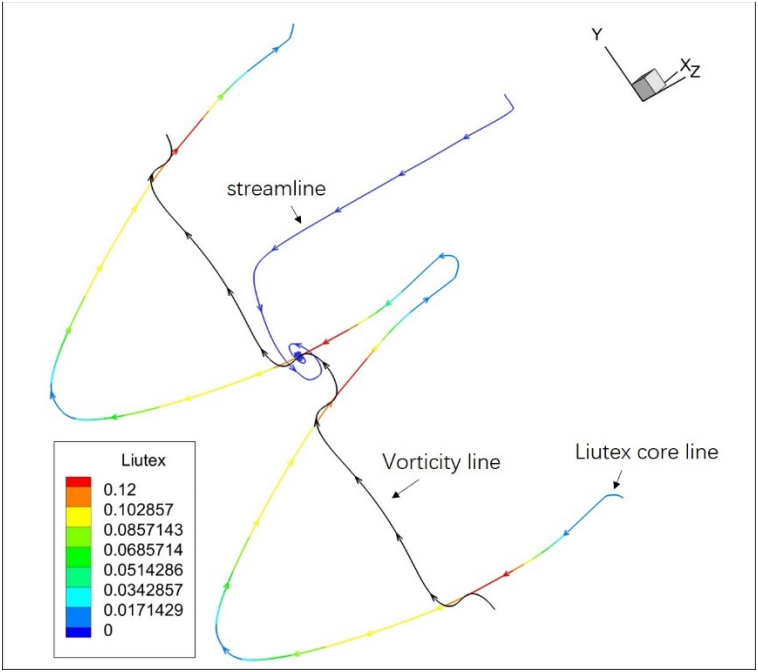

Fig. 2 Liutex core line, vorticity line and streamline

In general, their work is linked to "vorticity dynamics" and has nothing to do with "vortex dynamics" since vortex is not correlated to vorticity. On the other hand, all eigenvalue-based vortex identification methods are scalar leading to strongly threshold-dependent and seriously contaminated by shears and/or stretch (compression) as all elements of the velocity gradient tensor make contributions to the vortex identification criteria. This issue prompts Kolář (2007) [7] to formulate a triple decomposition from which the residual vorticity can be obtained after the extraction of an effective pure shearing motion and represents a direct and accurate measure of the pure rigid-body rotation of a fluid element. However, the triple decomposition requires a basic reference frame (BRF) to be first determined. Searching for the basic reference frame in 3D cases will result in an expensive optimization problem for every point in the flow field, which limits the applicability of the method (Epps, 2017) [17]. Kolář et al. (2013) [18] also introduced the concepts of the maximum corotation and the average corotation of line segments near a point and apply these methods for vortex identification. However, the maximum-corotation method suffers from that the maxima and the averaged corotation vector is evaluated by integration over a unit sphere, which makes it difficult to be used to study the transport property of vortex. Anyway, perfect solution for the optimization has not been found yet.

The vector and tensor decomposition should be unique and have clear physical meaning, like rotation, stretch (compression) and shear. This must require a unique coordinate and unique decomposition, which have been introduced in our previous work and called "Principal Coordinate" and "Principal Decomposition" of vorticity and velocity gradient tensor (Yu et al., 2020 [6]). These decompositions are not only unique but also Galilean invariant. The principal decomposition, which is based on Liutex vector and Liutex tensor, is unique for correct vorticity and tensor decomposition. They should replace the traditional Helmholtz decomposition and Cauchy-Stokes tensor decomposition which have caused countless confusion in fluid dynamics.

Liutex is a new quantified mathematical definition for fluid rotation or vortex introduced by Liu et al. [3-5], which innovatively defines its direction as the local rotation axis and defines Liutex strength as twice the angular speed of the rigid rotation without shear contamination. Liutex is proved to be unique and

Galilean invariant [19]. After defining Liutex, many Liutex-related vortex identification methods in the field of fluid dynamics have been developed and it gradually forms the Liutex theoretical system [20]. Right now, this system includes Liutex similarity [21], Liutex core line [22], [23], Liutex-Omega method [24],[25], Objective Liutex [26], principal coordinated and principal decomposition [6, 27]. The principal coordinate is a unique coordinate under which rotation, shear, and stretching can be easily and correctly decomposed. Principal decomposition is the decomposition under the principal coordinate that correctly decomposes the velocity gradient tensor into the rotation, stretching and shear parts. Based on Liutex theorem, many applications have been achieved successfully. The explicit Liutex magnitude was given by Wang et al (2019) [28]. An experimental test of the correctness of Liutex was done by Guo et al. [29]. Liutex has a wide application prospect. Wang et al. (2021) [30] found that Liutex is suitable for the characterization and evolution analysis of vortical structures in hydro-energy machinery. Compared with the traditional vorticity method, Liutex can effectively identify the vortices with the origin in rigid rotation motions. Wang et al (2021) [31] use Liutex to study fluid transport equations and found that compared with the vorticity transport equation, Liutex transport equation can extract the intuitive vortex evolution characteristics, which is beneficial to guide the analysis and control of vortical flows in fluids engineering. Yan and Xie (2021) [32] examined the vortex structure of a jet in a restricted transonic cross-flow. They described for the first time the details of the vortices in this flow field and found a new vortex structure, named the longitudinal shear vortex. Zhao et al. (2020) [33] use Liutex for vortex identification in marine hydrodynamics and Zhao et al. (2021) [34] apply Liutex to develop force field models in vortex and cavitation control. In Ref [35], Zhou et al. elaborated the hydrodynamic instabilities induced turbulent mixing in wide areas including inertial confinement fusion, supernovae and their transition criteria. Zhou detailed described Rayleigh–Taylor and Richtmyer–Meshkov instability and some related models in Ref. [36],[37]. Zhou's analysis is systematical, comprehensive, and sophisticated for flow instability and vortex generation, covering long history and state-of-the-art advances in turbulence research, which has clearly shown guidance for further and deeper scientific research. Another topic of general interest is turbulence mechanism. Xia et al. [38],[39] did many deep theory and experiment research on this topic, especially the

Rayleigh–Bénard (RB) convection. Liutex has the potential to help clarify the mechanism of these hydrodynamic instabilities or turbulence generation as Liutex is a vector definition which more accurately represents flow rotation or vortex in comparison with Q [40], $\Delta$ [41], $\lambda_{ci}$ [42] or $\lambda_2$ [43] methods which are scalar and shear-contaminated.

Although the Liutex-based principal decomposition has clear physical meaning such as rigid rotation, pure shear, pure stretching, the principal decomposition was made in a Liutex-based principal coordinate. Each point has its own coordinate which is different from its neighboring points. This makes the original principal decomposition meaningful only in theory but not in practical applications. The current work derived the decomposition in the original xyz coordinates. In this way, the new principal velocity tensor decomposition could become a practical tool for flow decomposition to replace the traditional Helmholtz velocity or Cauchy-Stokes tensor decomposition and provides a foundation for further development of fluid dynamics, especially for vortex science and turbulence research.

The paper is organized as follows. Section 1 gives a short introduction about the development in velocity gradient tensor decomposition. In section 2, the principal coordinate and principal decomposition are briefly reviewed. Section 3 provides the detailed principal decomposition of the velocity gradient tensor in the original xyz- system, which makes the principal decomposition viable as the new fluid kinematics. In Section 4, mathematical check of the new principal decomposition is made. Section 5 provides some numerical examples to show the principal decomposition in a boundary layer transition. Some conclusions are made in the concluding section.

## 2. Principal coordinates and principal decomposition of velocity gradient

### 2.1 Cauchy-Stokes decomposition in the xyz coordinate system

In the xyz coordinates, the Cauchy-Stokes decomposition gives the following formula:

$$\nabla \vec{v} = \begin{bmatrix} \frac{\partial u}{\partial x} & \frac{\partial u}{\partial y} & \frac{\partial u}{\partial z} \\ \frac{\partial v}{\partial x} & \frac{\partial v}{\partial y} & \frac{\partial v}{\partial z} \\ \frac{\partial w}{\partial x} & \frac{\partial w}{\partial y} & \frac{\partial w}{\partial z} \end{bmatrix} = A + B, \tag{2.1}$$

where $A$ and $B$ are the symmetric (strain-rate tensor) part and anti-symmetric (vorticity tensor) part of the velocity gradient tensor.

$$A = \frac{1}{2}(\nabla \vec{v} + \nabla \vec{v}^T) = \begin{bmatrix} \frac{\partial u}{\partial x} & \frac{1}{2}\left(\frac{\partial u}{\partial y} + \frac{\partial v}{\partial x}\right) & \frac{1}{2}\left(\frac{\partial u}{\partial z} + \frac{\partial w}{\partial x}\right) \\ \frac{1}{2}\left(\frac{\partial v}{\partial x} + \frac{\partial u}{\partial y}\right) & \frac{\partial v}{\partial y} & \frac{1}{2}\left(\frac{\partial v}{\partial z} + \frac{\partial w}{\partial y}\right) \\ \frac{1}{2}\left(\frac{\partial w}{\partial x} + \frac{\partial u}{\partial z}\right) & \frac{1}{2}\left(\frac{\partial w}{\partial y} + \frac{\partial v}{\partial z}\right) & \frac{\partial w}{\partial z} \end{bmatrix} \quad (2.2)$$

$$B = \frac{1}{2}(\nabla \vec{v} - \nabla \vec{v}^T) = \begin{bmatrix} 0 & \frac{1}{2}\left(\frac{\partial u}{\partial y} - \frac{\partial v}{\partial x}\right) & \frac{1}{2}\left(\frac{\partial u}{\partial z} - \frac{\partial w}{\partial x}\right) \\ \frac{1}{2}\left(\frac{\partial v}{\partial x} - \frac{\partial u}{\partial y}\right) & 0 & \frac{1}{2}\left(\frac{\partial v}{\partial z} - \frac{\partial w}{\partial y}\right) \\ \frac{1}{2}\left(\frac{\partial w}{\partial x} - \frac{\partial u}{\partial z}\right) & \frac{1}{2}\left(\frac{\partial w}{\partial y} - \frac{\partial v}{\partial z}\right) & 0 \end{bmatrix} \quad (2.3)$$

## 2.2 Liutex and Principal coordinates

***Definition 1***: Liutex is a vector defined as

$$\vec{R} = R\vec{r} \quad (2.4)$$

where $R$ is the magnitude of Liutex and $\vec{r}$ is the local rotation axis.

According to Gao and Liu [4], $\vec{r}$ is the real eigenvector of the velocity gradient $\nabla \vec{v}$

Wang et al. [28] proposed an explicit expression of $R$ i.e.

$$R = \vec{\omega} \cdot \vec{r} - \sqrt{(\vec{\omega} \cdot \vec{r})^2 - 4\lambda_{ci}^2} \quad (2.5)$$

where $\vec{\omega}$ is vorticity, $\vec{r}$ is the local rotation axis, and $\lambda_{ci}$ is the imaginary part of $\nabla \vec{v}$ complex eigenvalue.

In order to make $\vec{r}$ unique, it is required that $\vec{\omega} \cdot \vec{r} > 0$

***Definition 2***: Principal coordinate at a point is a coordinate that satisfies:

(1) Its Z-axis is parallel to the $\vec{r}$ (direction of Liutex [4])
(2) The velocity gradient tensor under this coordinate is in the form of:

$$\begin{bmatrix} \lambda_{cr} & \frac{\partial U}{\partial Y} & 0 \\ \frac{\partial V}{\partial X} & \lambda_{cr} & 0 \\ \frac{\partial W}{\partial X} & \frac{\partial W}{\partial Y} & \lambda_r \end{bmatrix}, \quad (2.6)$$

where $\lambda_r$, $\lambda_{cr}$ are real eigenvalue and real part of the conjugate complex eigenvalue pair of the velocity gradient tensor respectively for rotation points.

(3) $\frac{\partial U}{\partial Y} < 0$

### 2.3 Principal decomposition of velocity gradient tensor in principal coordinates

In the principal coordinates, the velocity gradient tensor can be written as

$$\nabla \vec{V} = \begin{bmatrix} \lambda_{cr} & -R/2 & 0 \\ R/2 + \epsilon & \lambda_{cr} & 0 \\ \xi & \eta & \lambda_r \end{bmatrix} \qquad (2.7)$$

#### 2.3.1 Cauchy-Stokes decomposition in principal coordinate system

$$\nabla \vec{V} = \begin{bmatrix} \lambda_{cr} & -\frac{R}{2} & 0 \\ \frac{R}{2} + \epsilon & \lambda_{cr} & 0 \\ \xi & \eta & \lambda_r \end{bmatrix} = A + B \qquad (2.8)$$

$$A = \begin{bmatrix} \lambda_{cr} & \frac{\epsilon}{2} & \frac{\xi}{2} \\ \frac{\epsilon}{2} & \lambda_{cr} & \frac{\eta}{2} \\ \frac{\xi}{2} & \frac{\eta}{2} & \lambda_r \end{bmatrix} \qquad (2.9)$$

$$B = \begin{bmatrix} 0 & -\frac{R}{2} - \frac{\epsilon}{2} & -\frac{\xi}{2} \\ \frac{R}{2} + \frac{\epsilon}{2} & 0 & -\frac{\eta}{2} \\ \frac{\xi}{2} & \frac{\eta}{2} & 0 \end{bmatrix} = \begin{bmatrix} 0 & -\frac{1}{2}\omega_Z & \frac{1}{2}\omega_X \\ \frac{1}{2}\omega_Z & 0 & -\frac{1}{2}\omega_Y \\ -\frac{1}{2}\omega_X & \frac{1}{2}\omega_Y & 0 \end{bmatrix} \text{ where } \vec{\omega} \text{ is the vorticity} \qquad (2.10)$$

Apparently, $\xi, \eta, \epsilon$ can be determined by vorticity and Liutex: $\xi = \omega_X, \eta = \omega_Y, \epsilon = \omega_Z - R$

#### 2.3.2 UTA R-NR decomposition in principal coordinate

The **UTA R-NR** decomposition is to decompose the velocity gradient tensor to a rigid rotational part and a non-rotational part

$$\nabla \vec{V} = R + NR \qquad (2.11)$$

$$R = \begin{bmatrix} 0 & -R/2 & 0 \\ R/2 & 0 & 0 \\ 0 & 0 & 0 \end{bmatrix}$$

$$NR = \begin{bmatrix} \lambda_{cr} & 0 & 0 \\ s & \lambda_{cr} & 0 \\ \xi & \eta & \lambda_r \end{bmatrix}$$

### 2.3.3 Vorticity decomposition in principal coordinates

$$B = \begin{bmatrix} 0 & -\frac{R}{2}-\frac{\epsilon}{2} & -\frac{\xi}{2} \\ \frac{R}{2}+\frac{\epsilon}{2} & 0 & -\frac{\eta}{2} \\ \frac{\xi}{2} & \frac{\eta}{2} & 0 \end{bmatrix} = \begin{bmatrix} 0 & -\frac{R}{2} & 0 \\ \frac{R}{2} & 0 & 0 \\ 0 & 0 & 0 \end{bmatrix} = \begin{bmatrix} 0 & -\frac{\epsilon}{2} & -\frac{\xi}{2} \\ \frac{\epsilon}{2} & 0 & -\frac{\eta}{2} \\ \frac{\xi}{2} & \frac{\eta}{2} & 0 \end{bmatrix} = R + VS, \qquad (2.12)$$

where $R$ is Liutex tensor and $VS$ is called vorticity shear following Kolar (2007)

### 2.3.4 Principal tensor decomposition in the principal coordinates

$$\nabla \vec{V} = \begin{bmatrix} \lambda_{cr} & -\frac{R}{2} & 0 \\ \frac{R}{2}+\epsilon & \lambda_{cr} & 0 \\ \xi & \eta & \lambda_r \end{bmatrix} = \begin{bmatrix} 0 & -R/2 & 0 \\ R/2 & 0 & 0 \\ 0 & 0 & 0 \end{bmatrix} + \begin{bmatrix} \lambda_{cr} & 0 & 0 \\ 0 & \lambda_{cr} & 0 \\ 0 & 0 & \lambda_r \end{bmatrix} + \begin{bmatrix} 0 & 0 & 0 \\ \epsilon & 0 & 0 \\ \xi & \eta & 0 \end{bmatrix} = R + SC + S \quad (2.13)$$

where $R$ is rigid rotation, $SC$ is stretching (compression), $S$ is the shear, and $\xi = \omega_X, \eta = \omega_Y, \epsilon = \omega_Z - R$.

Being different from Cauchy-Stokes decomposition, the principal decomposition is Galilean invariant, i.e. independent of coordinate rotation or translation as all elements in the three matrices are Galilean invariant note that vorticity, Liutex and eigenvalues are all Galilean invariant.

According to Kolar (2007), a symmetric shear ($SS$) is defined as

$$SS = \begin{bmatrix} 0 & \frac{\epsilon}{2} & \frac{\xi}{2} \\ \frac{\epsilon}{2} & 0 & \frac{\eta}{2} \\ \frac{\xi}{2} & \frac{\eta}{2} & 0 \end{bmatrix} \qquad (2.14)$$

The shear is a summation of vorticity shear plus the strain (symmetric) shear

$$S = VS + SS = \begin{bmatrix} 0 & \frac{\epsilon}{2} & \frac{\xi}{2} \\ \frac{\epsilon}{2} & 0 & \frac{\eta}{2} \\ \frac{\xi}{2} & \frac{\eta}{2} & 0 \end{bmatrix} + \begin{bmatrix} 0 & -\frac{\epsilon}{2} & -\frac{\xi}{2} \\ \frac{\epsilon}{2} & 0 & -\frac{\eta}{2} \\ \frac{\xi}{2} & \frac{\eta}{2} & 0 \end{bmatrix} = \begin{bmatrix} 0 & 0 & 0 \\ \epsilon & 0 & 0 \\ \xi & \eta & 0 \end{bmatrix} \qquad (2.15)$$

### 3. Principal decomposition in the original xyz- coordinates

All points in a flow field can be classified as rotational or irrotational points. In this section, the expression for rotational points will be presented first, and then the expression for irrotational points is followed.

### 3.1 Galilean invariant

Physical quantity which is independent of coordinate translation or rotation is called Galilean invariant. Vorticity, Liutex, eigenvalues of $\nabla \vec{V}$ are all Galilean invariant or unchanged in different coordinates.

### 3.2 *UTA R-NR* decomposition in the xyz coordinate

In the original xyz coordinate, the velocity gradient tensor is

$$\nabla \vec{v} = \begin{bmatrix} \frac{\partial u}{\partial x} & \frac{\partial u}{\partial y} & \frac{\partial u}{\partial z} \\ \frac{\partial v}{\partial x} & \frac{\partial v}{\partial y} & \frac{\partial v}{\partial z} \\ \frac{\partial w}{\partial x} & \frac{\partial w}{\partial y} & \frac{\partial w}{\partial z} \end{bmatrix} = R + NR \tag{3.1}$$

Since Liutex $\vec{R}$ is Galilean invariant, its components in the xyz coordinate are $R_x, R_y, R_z$ respectively. The corresponding rotational tensor is

$$R = \frac{1}{2}\begin{bmatrix} 0 & -R_z & R_y \\ R_z & 0 & -R_x \\ -R_y & R_x & 0 \end{bmatrix} \tag{3.2}$$

As Liutex vector has been obtained, $R_x, R_y, R_z$ are easily calculated, and the **UTA R-NR** decomposition is straightforward in the original xyz coordinate,

$$\nabla \vec{v} = \begin{bmatrix} \frac{\partial u}{\partial x} & \frac{\partial u}{\partial y} & \frac{\partial u}{\partial z} \\ \frac{\partial v}{\partial x} & \frac{\partial v}{\partial y} & \frac{\partial v}{\partial z} \\ \frac{\partial w}{\partial x} & \frac{\partial w}{\partial y} & \frac{\partial w}{\partial z} \end{bmatrix} = \frac{1}{2}\begin{bmatrix} 0 & -R_z & R_y \\ R_z & 0 & -R_x \\ -R_y & R_x & 0 \end{bmatrix} + \begin{bmatrix} \frac{\partial u}{\partial x} & \frac{\partial u}{\partial y}+\frac{1}{2}R_z & \frac{\partial u}{\partial z}-\frac{1}{2}R_y \\ \frac{\partial v}{\partial x}-\frac{1}{2}R_z & \frac{\partial v}{\partial y} & \frac{\partial v}{\partial z}+\frac{1}{2}R_x \\ \frac{\partial w}{\partial x}+\frac{1}{2}R_y & \frac{\partial w}{\partial y}-\frac{1}{2}R_x & \frac{\partial w}{\partial z} \end{bmatrix} = R + NR \tag{3.3}$$

For the discussion below, it is assumed that Liutex is non-zero at the points which means there are rotations.

### 3.2 Transform from Principal coordinate back to the original xyz coordinates

The relation of velocity gradient tensor in principal coordinates and in the original xyz coordinates (Liu et al, 2018) [3] is:

$$\nabla \vec{V} = P^T Q^T \nabla \vec{v} Q P \quad \text{or} \tag{3.4}$$

$$\nabla \vec{v} = Q P \nabla \vec{V} P^T Q^T \tag{3.5}$$

where $Q$ and $P$ are orthogonal grid-rotation matrix

According to principal decomposition, $\nabla \vec{V} = R + VS + SS + SC = R + S + SC$, where

**R** is the rigid rotation, **VS** is the vorticity (anti-symmetric) shear, **SS** is the strain (symmetric) shear and **SC** is the stretching (compression). The rotation back to the original xyz- system gives

$$\nabla \vec{v} = QP(R)P^T Q^T + QP(VS)P^T Q^T + QP(SS)P^T Q^T + QP(SC)P^T Q^T = \widetilde{R} + \widetilde{VS} + \widetilde{SS} + \widetilde{SC} \tag{3.6}$$

$$\nabla \vec{v} = QP(R)P^T Q^T + QP(S)P^T Q^T + QP(SC)P^T Q^T = \widetilde{R} + \widetilde{S} + \widetilde{SC} \tag{3.7}$$

However, this transformation involves in two coordinate rotation, i.e. $P$ and $Q$ rotations and, therefore, is inconvenient. The following work tries to get the decomposition directly by using Liutex, vorticity, eigenvalues and eigenvectors which are Galilean invariant.

## 3.3 The principal decomposition in the xyz coordinate

As addressed above, the principal decomposition in the original xyz coordinates can be obtained by inverse of $Q$ and $P$ rotation:

$$\nabla \vec{v} = QP(R)P^TQ^T + QP(S)P^TQ^T + QP(SC)P^TQ^T = \tilde{R} + \tilde{S} + \widetilde{SC} \tag{3.8}$$

For simplification, $Q$ and $P$ rotation can be combined as one orthogonal rotation matrix, say $Q$. From now, the principal decomposition in the xyz coordinate is written as follows.

$$\nabla \vec{v} = Q(R)Q^T + Q(S)Q^T + Q(SC)Q^T = \tilde{R} + \tilde{S} + \widetilde{SC} \tag{3.9}$$

Here, $Q$ is equivalent as $QP$ which was used in the previous sections.

### 3.3.1 Rotation matrix $Q$ in xyz coordinate

**Definition 3.** $Q^T$ is defined as a rotation matrix to rotate the z-axis to parallel to $\vec{r}$, where

$$Q = \begin{bmatrix} Q_{11} & Q_{12} & Q_{13} \\ Q_{21} & Q_{22} & Q_{23} \\ Q_{31} & Q_{32} & Q_{33} \end{bmatrix}, \tag{3.10}$$

**Theorem 1.** If the third column of rotation $Q$ is $\vec{r}$, $Q^T$ can rotate the z-axis to parallel to $\vec{r}$

Proof. If $Q = \begin{bmatrix} Q_{11} & Q_{12} & r_x \\ Q_{21} & Q_{22} & r_y \\ Q_{31} & Q_{32} & r_z \end{bmatrix}$ \hfill (3.11)

$$Q^T \begin{bmatrix} r_x \\ r_y \\ r_z \end{bmatrix} = \begin{bmatrix} Q_{11} & Q_{12} & Q_{13} \\ Q_{21} & Q_{22} & Q_{23} \\ r_x & r_y & r_z \end{bmatrix} \begin{bmatrix} r_x \\ r_y \\ r_z \end{bmatrix} = \begin{bmatrix} 0 \\ 0 \\ 1 \end{bmatrix},$$

where $r_x, r_y, r_z$ are unit Liutex (real eigenvector of $\nabla \vec{v}$) components in the xyz coordinate respectively and has been obtained. For rotational points, Liutex exists, so $r_x, r_y, r_z$ must exist. Note that $Q$ is an orthogonal matrix so that columns are orthogonal to each other.

It will be seen that other elements will not be used in the principal decomposition except for $r_x, r_y, r_z$

### 3.3.2 Liutex part in the xyz system

Liutex R is Galilean invariant and is therefore easy to write in the xyz coordinates

In the principal coordinates, $\boldsymbol{R} = \begin{bmatrix} 0 & -\frac{R}{2} & 0 \\ \frac{R}{2} & 0 & 0 \\ 0 & 0 & 0 \end{bmatrix}$

In the original xyz coordinates, $\widetilde{\boldsymbol{R}} = \boldsymbol{Q}(\boldsymbol{R})\boldsymbol{Q}^T = \begin{bmatrix} 0 & -\frac{R_z}{2} & \frac{R_y}{2} \\ \frac{R_z}{2} & 0 & -\frac{R_x}{2} \\ -\frac{R_y}{2} & \frac{R_x}{2} & 0 \end{bmatrix}$ (3.12)

where $R = \sqrt{R_x^2 + R_y^2 + R_z^2}$

### 3.3.3 Stretching (Compression) part $SC$ in the xyz system

The stretching (compression) part $\boldsymbol{SC}$ in the principal coordinates can be written as

$$\boldsymbol{SC} = \begin{bmatrix} \lambda_{cr} & 0 & 0 \\ 0 & \lambda_{cr} & 0 \\ 0 & 0 & \lambda_r \end{bmatrix} \quad (3.13)$$

In the original xyz coordinate, the stretching (compression) part can be obtained by anti-$\boldsymbol{Q}$ rotation

$$\widetilde{\boldsymbol{SC}} = \boldsymbol{Q}(\boldsymbol{SC})\boldsymbol{Q}^T = \boldsymbol{Q}\begin{bmatrix} \lambda_{cr} & 0 & 0 \\ 0 & \lambda_{cr} & 0 \\ 0 & 0 & \lambda_r \end{bmatrix}\boldsymbol{Q}^T = \boldsymbol{Q}\begin{bmatrix} \lambda_{cr} & 0 & 0 \\ 0 & \lambda_{cr} & 0 \\ 0 & 0 & \lambda_{cr} \end{bmatrix}\boldsymbol{Q}^T + \boldsymbol{Q}\begin{bmatrix} 0 & 0 & 0 \\ 0 & 0 & 0 \\ 0 & 0 & \lambda_r - \lambda_{cr} \end{bmatrix}\boldsymbol{Q}^T \quad (3.14)$$

Since $\boldsymbol{Q}$ is orthogonal and $\boldsymbol{Q}\boldsymbol{Q}^T = \boldsymbol{I}$, $\widetilde{\boldsymbol{SC}} = \begin{bmatrix} \lambda_{cr} & 0 & 0 \\ 0 & \lambda_{cr} & 0 \\ 0 & 0 & \lambda_{cr} \end{bmatrix} + \boldsymbol{Q}\begin{bmatrix} 0 & 0 & 0 \\ 0 & 0 & 0 \\ 0 & 0 & \lambda_r - \lambda_{cr} \end{bmatrix}\boldsymbol{Q}^T$ (3.15)

However, Q has the form $Q = \begin{bmatrix} Q_{11} & Q_{12} & r_x \\ Q_{21} & Q_{22} & r_y \\ Q_{31} & Q_{32} & r_z \end{bmatrix}$

$$\widetilde{\boldsymbol{SC}} = \begin{bmatrix} \lambda_{cr} & 0 & 0 \\ 0 & \lambda_{cr} & 0 \\ 0 & 0 & \lambda_{cr} \end{bmatrix} + \begin{bmatrix} Q_{11} & Q_{12} & r_x \\ Q_{21} & Q_{22} & r_y \\ Q_{31} & Q_{32} & r_z \end{bmatrix}\begin{bmatrix} 0 & 0 & 0 \\ 0 & 0 & 0 \\ 0 & 0 & \lambda_r - \lambda_{cr} \end{bmatrix}\begin{bmatrix} Q_{11} & Q_{21} & Q_{31} \\ Q_{12} & Q_{22} & Q_{32} \\ r_x & r_y & r_z \end{bmatrix}$$

$$\widetilde{\boldsymbol{SC}} = \begin{bmatrix} \lambda_{cr} & 0 & 0 \\ 0 & \lambda_{cr} & 0 \\ 0 & 0 & \lambda_{cr} \end{bmatrix} + (\lambda_r - \lambda_{cr})\begin{bmatrix} Q_{11} & Q_{12} & r_x \\ Q_{21} & Q_{22} & r_y \\ Q_{31} & Q_{32} & r_z \end{bmatrix}\begin{bmatrix} 0 & 0 & 0 \\ 0 & 0 & 0 \\ r_x & r_y & r_z \end{bmatrix}$$

$$\widetilde{\boldsymbol{SC}} = \begin{bmatrix} \lambda_{cr} & 0 & 0 \\ 0 & \lambda_{cr} & 0 \\ 0 & 0 & \lambda_{cr} \end{bmatrix} + (\lambda_r - \lambda_{cr})\begin{bmatrix} r_x^2 & r_x r_y & r_x r_z \\ r_y r_x & r_y^2 & r_y r_z \\ r_z r_x & r_z r_y & r_z^2 \end{bmatrix} \quad (3.16)$$

### 3.3.4 Principal decomposition of velocity gradient in the xyz coordinates

**Theorem 2. The velocity gradient tensor can be decomposed as rigid rotation, pure shear and stretching (compression)**

Proof. $\nabla \vec{v} = Q(R)Q^T + Q(S)Q^T + Q(SC)Q^T = \tilde{R} + \tilde{S} + \widetilde{SC}$

$$\begin{bmatrix} \frac{\partial u}{\partial x} & \frac{\partial u}{\partial y} & \frac{\partial u}{\partial z} \\ \frac{\partial v}{\partial x} & \frac{\partial v}{\partial y} & \frac{\partial v}{\partial z} \\ \frac{\partial w}{\partial x} & \frac{\partial w}{\partial y} & \frac{\partial w}{\partial z} \end{bmatrix} = \frac{1}{2}\begin{bmatrix} 0 & -R_z & R_y \\ R_z & 0 & -R_x \\ -R_y & R_x & 0 \end{bmatrix} + \tilde{S} + \begin{bmatrix} \lambda_{cr} & 0 & 0 \\ 0 & \lambda_{cr} & 0 \\ 0 & 0 & \lambda_{cr} \end{bmatrix} + (\lambda_r - \lambda_{cr})\begin{bmatrix} r_x^2 & r_x r_y & r_x r_z \\ r_y r_x & r_y^2 & r_y r_z \\ r_z r_x & r_z r_y & r_z^2 \end{bmatrix} \quad (3.17)$$

The shear part of the velocity gradient tensor is

$$\tilde{S} = \begin{bmatrix} \frac{\partial u}{\partial x} & \frac{\partial u}{\partial y} & \frac{\partial u}{\partial z} \\ \frac{\partial v}{\partial x} & \frac{\partial v}{\partial y} & \frac{\partial v}{\partial z} \\ \frac{\partial w}{\partial x} & \frac{\partial w}{\partial y} & \frac{\partial w}{\partial z} \end{bmatrix} - \frac{1}{2}\begin{bmatrix} 0 & -R_z & R_y \\ R_z & 0 & -R_x \\ -R_y & R_x & 0 \end{bmatrix} - \begin{bmatrix} \lambda_{cr} & 0 & 0 \\ 0 & \lambda_{cr} & 0 \\ 0 & 0 & \lambda_{cr} \end{bmatrix} - (\lambda_r - \lambda_{cr})\begin{bmatrix} r_x^2 & r_x r_y & r_x r_z \\ r_y r_x & r_y^2 & r_y r_z \\ r_z r_x & r_z r_y & r_z^2 \end{bmatrix} \quad (3.18)$$

$$\tilde{S} = \begin{bmatrix} \frac{\partial u}{\partial x} & \frac{\partial u}{\partial y} + R_z/2 & \frac{\partial u}{\partial z} - R_y/2 \\ \frac{\partial v}{\partial x} - R_z/2 & \frac{\partial v}{\partial y} & \frac{\partial v}{\partial z} + R_x/2 \\ \frac{\partial w}{\partial x} + R_y/2 & \frac{\partial w}{\partial y} - R_x/2 & \frac{\partial w}{\partial z} \end{bmatrix} - \begin{bmatrix} \lambda_{cr} + (\lambda_r - \lambda_{cr})r_x^2 & (\lambda_r - \lambda_{cr})r_x r_y & (\lambda_r - \lambda_{cr})r_x r_z \\ (\lambda_r - \lambda_{cr})r_y r_x & \lambda_{cr} + (\lambda_r - \lambda_{cr})r_y^2 & (\lambda_r - \lambda_{cr})r_y r_z \\ (\lambda_r - \lambda_{cr})r_z r_x & (\lambda_r - \lambda_{cr})r_z r_y & \lambda_{cr} + (\lambda_r - \lambda_{cr})r_z^2 \end{bmatrix} \quad (3.19)$$

Next, those points where exist no rotations are discussed. Since there is no rotation, the velocity gradient tensor has three real eigenvalues.

### 3.3.5 Principal decomposition of the velocity gradient in the xyz coordinates (irrotational points)

Based on Schur decomposition, if $\nabla \vec{v}$ is real there exists orthogonal matrix Q such that

$$\nabla \vec{V} = Q \nabla \vec{v} Q^{-1} = \begin{bmatrix} \lambda_1 & 0 & 0 \\ \epsilon & \lambda_2 & 0 \\ \xi & \eta & \lambda_3 \end{bmatrix} \quad (3.20)$$

where $\nabla \vec{V}$ is the velocity gradient tensor in the principal coordinate which is a lower triangle matrix and. $\lambda_1, \lambda_2$ and $\lambda_3$ are the three real eigenvalues of the $\nabla \vec{V}$ or $\nabla \vec{v}$. Q is the transformation matrix from original coordinate to the principal coordinates.

$\nabla \vec{V}$ can be decomposed into rigid rotation, pure shear and stretching parts as follow

$$\nabla \vec{V} = \begin{bmatrix} 0 & 0 & 0 \\ 0 & 0 & 0 \\ 0 & 0 & 0 \end{bmatrix} + \begin{bmatrix} \lambda_1 & 0 & 0 \\ 0 & \lambda_2 & 0 \\ 0 & 0 & \lambda_3 \end{bmatrix} + \begin{bmatrix} 0 & 0 & 0 \\ \epsilon & 0 & 0 \\ \xi & \eta & 0 \end{bmatrix} = R + SC + S \qquad (3.21)$$

where R, SC and S represent rotation, stretching and shear matrices respectively. R is a zero matrix because there is no rotation at the point.

Next step is to rotate R, SC and S back to the original xyz coordinate. Suppose the transformation matrix Q is

$$Q = \begin{bmatrix} Q_{11} & Q_{12} & Q_{13} \\ Q_{21} & Q_{22} & Q_{23} \\ Q_{31} & Q_{32} & Q_{33} \end{bmatrix} \qquad (3.22)$$

Schur decomposition is a classical theory and it is easy for people to find ready-made program to calculate Q. Let $\widetilde{R}$, $\widetilde{SC}$, $\widetilde{S}$ be the rotation, stretching and shear matrix in the original coordinate. Since there is no rotation, $\widetilde{R}$ is a zero matrix.

$$\widetilde{R} = \begin{bmatrix} 0 & 0 & 0 \\ 0 & 0 & 0 \\ 0 & 0 & 0 \end{bmatrix} \qquad (3.23)$$

And,

$$\widetilde{SC} = Q(SC)Q^{-1} \qquad (3.24)$$

where SC is the stretching matrix in the principal coordinate.

$$\begin{aligned}
\widetilde{SC} &= \begin{bmatrix} Q_{11} & Q_{12} & Q_{13} \\ Q_{21} & Q_{22} & Q_{23} \\ Q_{31} & Q_{32} & Q_{33} \end{bmatrix} \begin{bmatrix} \lambda_1 & 0 & 0 \\ 0 & \lambda_2 & 0 \\ 0 & 0 & \lambda_3 \end{bmatrix} \begin{bmatrix} Q_{11} & Q_{21} & Q_{31} \\ Q_{12} & Q_{22} & Q_{32} \\ Q_{13} & Q_{23} & Q_{33} \end{bmatrix} \\
&= Q \begin{bmatrix} \lambda_1 & 0 & 0 \\ 0 & 0 & 0 \\ 0 & 0 & 0 \end{bmatrix} Q^{-1} + Q \begin{bmatrix} 0 & 0 & 0 \\ 0 & \lambda_2 & 0 \\ 0 & 0 & 0 \end{bmatrix} Q^{-1} + Q \begin{bmatrix} 0 & 0 & 0 \\ 0 & 0 & 0 \\ 0 & 0 & \lambda_3 \end{bmatrix} Q^{-1} \\
&= \lambda_1 \begin{bmatrix} Q_{11}Q_{11} & Q_{21}Q_{11} & Q_{31}Q_{11} \\ Q_{11}Q_{21} & Q_{21}Q_{21} & Q_{31}Q_{21} \\ Q_{11}Q_{31} & Q_{21}Q_{31} & Q_{31}Q_{31} \end{bmatrix} + \lambda_2 \begin{bmatrix} Q_{12}Q_{12} & Q_{22}Q_{12} & Q_{32}Q_{12} \\ Q_{12}Q_{22} & Q_{22}Q_{22} & Q_{32}Q_{22} \\ Q_{12}Q_{32} & Q_{22}Q_{32} & Q_{32}Q_{32} \end{bmatrix} + \lambda_3 \begin{bmatrix} Q_{13}Q_{13} & Q_{23}Q_{13} & Q_{33}Q_{13} \\ Q_{13}Q_{23} & Q_{23}Q_{23} & Q_{33}Q_{23} \\ Q_{31}Q_{33} & Q_{23}Q_{33} & Q_{33}Q_{33} \end{bmatrix} \\
&= \begin{bmatrix} \lambda_1 Q_{11}Q_{11} + \lambda_2 Q_{12}Q_{12} + \lambda_3 Q_{13}Q_{13} & \lambda_1 Q_{21}Q_{11} + \lambda_2 Q_{22}Q_{12} + \lambda_3 Q_{23}Q_{13} & \lambda_1 Q_{31}Q_{11} + \lambda_2 Q_{32}Q_{12} + \lambda_3 Q_{33}Q_{13} \\ \lambda_1 Q_{11}Q_{21} + \lambda_2 Q_{12}Q_{22} + \lambda_3 Q_{13}Q_{23} & \lambda_1 Q_{21}Q_{21} + \lambda_2 Q_{22}Q_{22} + \lambda_3 Q_{23}Q_{23} & \lambda_1 Q_{31}Q_{21} + \lambda_2 Q_{32}Q_{22} + \lambda_3 Q_{33}Q_{23} \\ \lambda_1 Q_{11}Q_{31} + \lambda_2 Q_{12}Q_{32} + \lambda_3 Q_{31}Q_{33} & \lambda_1 Q_{21}Q_{31} + \lambda_2 Q_{22}Q_{32} + \lambda_3 Q_{23}Q_{33} & \lambda_1 Q_{31}Q_{31} + \lambda_2 Q_{32}Q_{32} + \lambda_3 Q_{33}Q_{33} \end{bmatrix}
\end{aligned}$$

(3.25)

Denote

$$\widetilde{SC} = \begin{bmatrix} (\widetilde{SC})_{11} & (\widetilde{SC})_{12} & (\widetilde{SC})_{13} \\ (\widetilde{SC})_{21} & (\widetilde{SC})_{22} & (\widetilde{SC})_{23} \\ (\widetilde{SC})_{31} & (\widetilde{SC})_{32} & (\widetilde{SC})_{33} \end{bmatrix} \qquad (3.26)$$

Then,

$$\tilde{S} = \nabla\vec{v} - \widetilde{SC} - \widetilde{R} = \begin{bmatrix} \frac{\partial u}{\partial x} - (\widetilde{SC})_{11} & \frac{\partial u}{\partial y} - (\widetilde{SC})_{12} & \frac{\partial u}{\partial z} - (\widetilde{SC})_{13} \\ \frac{\partial v}{\partial x} - (\widetilde{SC})_{21} & \frac{\partial v}{\partial y} - (\widetilde{SC})_{22} & \frac{\partial v}{\partial z} - (\widetilde{SC})_{23} \\ \frac{\partial w}{\partial x} - (\widetilde{SC})_{31} & \frac{\partial w}{\partial y} - (\widetilde{SC})_{32} & \frac{\partial w}{\partial z} - (\widetilde{SC})_{33} \end{bmatrix} \quad (3.27)$$

3.3.6 Summary of the expressions of the velocity gradient in xyz coordinate for rotational and irrotational points

For the rotational points,

$$\nabla\vec{v} = \widetilde{R} + \widetilde{SC} + \tilde{S} \quad (3.28)$$

where

$$\widetilde{R} = \begin{bmatrix} 0 & -\frac{R_z}{2} & \frac{R_y}{2} \\ \frac{R_z}{2} & 0 & -\frac{R_x}{2} \\ -\frac{R_y}{2} & \frac{R_x}{2} & 0 \end{bmatrix} \quad (3.29)$$

$$\widetilde{SC} = \begin{bmatrix} \lambda_{cr} & 0 & 0 \\ 0 & \lambda_{cr} & 0 \\ 0 & 0 & \lambda_{cr} \end{bmatrix} + (\lambda_r - \lambda_{cr}) \begin{bmatrix} r_x^2 & r_x r_y & r_x r_z \\ r_y r_x & r_y^2 & r_y r_z \\ r_z r_x & r_z r_y & r_z^2 \end{bmatrix} \quad (3.30)$$

$$\tilde{S} = \begin{bmatrix} \frac{\partial u}{\partial x} & \frac{\partial u}{\partial y} + R_z/2 & \frac{\partial u}{\partial z} - R_y/2 \\ \frac{\partial v}{\partial x} - R_z/2 & \frac{\partial v}{\partial y} & \frac{\partial v}{\partial z} + R_x/2 \\ \frac{\partial w}{\partial x} + R_y/2 & \frac{\partial w}{\partial y} - R_x/2 & \frac{\partial w}{\partial z} \end{bmatrix} - \begin{bmatrix} \lambda_{cr} + (\lambda_r - \lambda_{cr})r_x^2 & (\lambda_r - \lambda_{cr})r_x r_y & (\lambda_r - \lambda_{cr})r_x r_z \\ (\lambda_r - \lambda_{cr})r_y r_x & \lambda_{cr} + (\lambda_r - \lambda_{cr})r_y^2 & (\lambda_r - \lambda_{cr})r_y r_z \\ (\lambda_r - \lambda_{cr})r_z r_x & (\lambda_r - \lambda_{cr})r_z r_y & \lambda_{cr} + (\lambda_r - \lambda_{cr})r_z^2 \end{bmatrix} \quad (3.31)$$

For irrotational points,

$$\nabla\vec{v} = \widetilde{R} + \widetilde{SC} + \tilde{S} \quad (3.32)$$

where

$$\widetilde{R} = \begin{bmatrix} 0 & 0 & 0 \\ 0 & 0 & 0 \\ 0 & 0 & 0 \end{bmatrix} \quad (3.33)$$

$$\widetilde{SC} = \begin{bmatrix} \lambda_1 Q_{11} Q_{11} + \lambda_2 Q_{12} Q_{12} + \lambda_3 Q_{13} Q_{13} & \lambda_1 Q_{21} Q_{11} + \lambda_2 Q_{22} Q_{12} + \lambda_3 Q_{23} Q_{13} & \lambda_1 Q_{31} Q_{11} + \lambda_2 Q_{32} Q_{12} + \lambda_3 Q_{33} Q_{13} \\ \lambda_1 Q_{11} Q_{21} + \lambda_2 Q_{12} Q_{22} + \lambda_3 Q_{13} Q_{23} & \lambda_1 Q_{21} Q_{21} + \lambda_2 Q_{22} Q_{22} + \lambda_3 Q_{23} Q_{23} & \lambda_1 Q_{31} Q_{21} + \lambda_2 Q_{32} Q_{22} + \lambda_3 Q_{33} Q_{23} \\ \lambda_1 Q_{11} Q_{31} + \lambda_2 Q_{12} Q_{32} + \lambda_3 Q_{31} Q_{33} & \lambda_1 Q_{21} Q_{31} + \lambda_2 Q_{22} Q_{32} + \lambda_3 Q_{23} Q_{33} & \lambda_1 Q_{31} Q_{31} + \lambda_2 Q_{32} Q_{32} + \lambda_3 Q_{33} Q_{33} \end{bmatrix}$$

$$(3.34)$$

$$\tilde{S} = \nabla\vec{v} - \widetilde{SC} - \widetilde{R} = \begin{bmatrix} \frac{\partial u}{\partial x} - (\widetilde{SC})_{11} & \frac{\partial u}{\partial y} - (\widetilde{SC})_{12} & \frac{\partial u}{\partial z} - (\widetilde{SC})_{13} \\ \frac{\partial v}{\partial x} - (\widetilde{SC})_{21} & \frac{\partial v}{\partial y} - (\widetilde{SC})_{22} & \frac{\partial v}{\partial z} - (\widetilde{SC})_{23} \\ \frac{\partial w}{\partial x} - (\widetilde{SC})_{31} & \frac{\partial w}{\partial y} - (\widetilde{SC})_{32} & \frac{\partial w}{\partial z} - (\widetilde{SC})_{33} \end{bmatrix} \quad (3.35)$$

$\lambda_1$, $\lambda_2$ and $\lambda_3$ are the three real eigenvalues of velocity gradient tensor.

### 3.3.7  Vorticity shear and strain shear in the xyz coordinate

Since both vorticity and Liutex are obtained in the original xyz coordinate, it is easy to write the vorticity decomposition in the xyz coordinate

The vorticity tensor is

$$\widetilde{B} = \frac{1}{2}(\nabla \vec{v} - \nabla \vec{v}^T) = \begin{bmatrix} 0 & \frac{1}{2}\left(\frac{\partial u}{\partial y} - \frac{\partial v}{\partial x}\right) & \frac{1}{2}\left(\frac{\partial u}{\partial z} - \frac{\partial w}{\partial x}\right) \\ \frac{1}{2}\left(\frac{\partial v}{\partial x} - \frac{\partial u}{\partial y}\right) & 0 & \frac{1}{2}\left(\frac{\partial v}{\partial z} - \frac{\partial w}{\partial y}\right) \\ \frac{1}{2}\left(\frac{\partial w}{\partial x} - \frac{\partial u}{\partial z}\right) & \frac{1}{2}\left(\frac{\partial w}{\partial y} - \frac{\partial v}{\partial z}\right) & 0 \end{bmatrix} = \begin{bmatrix} 0 & -1/2\omega_z & 1/2\omega_y \\ 1/2\omega_z & 0 & -1/2\omega_x \\ -1/2\omega_y & 1/2\omega_x & 0 \end{bmatrix} \quad (3.36)$$

The Liutex tensor in the xyz system is

$$\widetilde{R} = \frac{1}{2}\begin{bmatrix} 0 & -R_z & R_y \\ R_z & 0 & -R_x \\ -R_y & R_x & 0 \end{bmatrix} \tag{3.37}$$

Since
$$\widetilde{B} = \widetilde{R} + \widetilde{VS}$$

$$\widetilde{VS} = \widetilde{B} - \widetilde{R} = \begin{bmatrix} 0 & -1/2(\omega_z - R_z) & 1/2(\omega_y - R_y) \\ 1/2(\omega_z - R_z) & 0 & -1/2(\omega_x - R_x) \\ -1/2(\omega_y - R_y) & 1/2(\omega_x - R_x) & 0 \end{bmatrix} \tag{3.38}$$

Consequently, in the xyz system, the vorticity decomposition gives $\nabla \times \vec{V} = \vec{R} + \vec{VS}$ or $\vec{VS} = \nabla \times \vec{V} - \vec{R}$ where $\vec{VS}$ is the vorticity shear. Fig. 3 shows the decomposition of vorticity in a direct numerical simulation result of boundary layer transition.

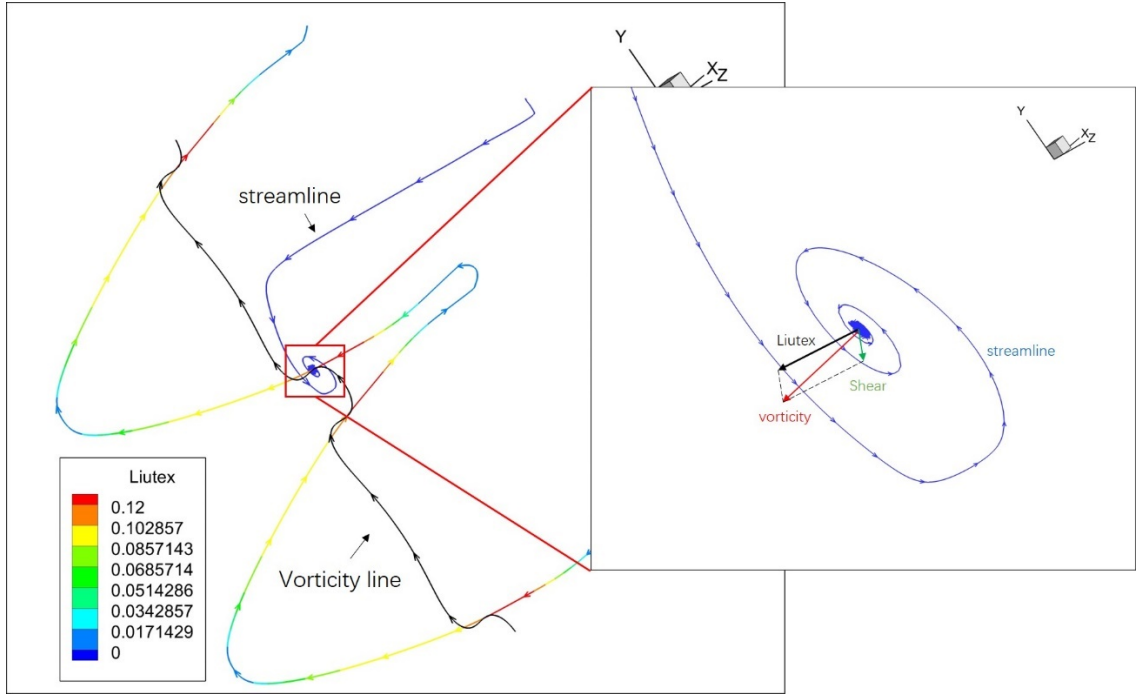

Fig. 3 Vorticity decomposition in a direct numerical simulation result of boundary layer transition.

The strain tensor is $\widetilde{A} = \frac{1}{2}(\nabla\vec{v} + \nabla\vec{v}^T) = \begin{bmatrix} \frac{\partial u}{\partial x} & \frac{1}{2}\left(\frac{\partial u}{\partial y} + \frac{\partial v}{\partial x}\right) & \frac{1}{2}\left(\frac{\partial u}{\partial z} + \frac{\partial w}{\partial x}\right) \\ \frac{1}{2}\left(\frac{\partial v}{\partial x} + \frac{\partial u}{\partial y}\right) & \frac{\partial v}{\partial y} & \frac{1}{2}\left(\frac{\partial v}{\partial z} + \frac{\partial w}{\partial y}\right) \\ \frac{1}{2}\left(\frac{\partial w}{\partial x} + \frac{\partial u}{\partial z}\right) & \frac{1}{2}\left(\frac{\partial w}{\partial y} + \frac{\partial v}{\partial z}\right) & \frac{\partial w}{\partial z} \end{bmatrix}$ (3.39)

and the stretching (compression) tensor is

$$\widetilde{SC} = \begin{bmatrix} \lambda_{cr} & 0 & 0 \\ 0 & \lambda_{cr} & 0 \\ 0 & 0 & \lambda_{cr} \end{bmatrix} + (\lambda_r - \lambda_{cr}) \begin{bmatrix} r_x^2 & r_x r_y & r_x r_z \\ r_y r_x & r_y^2 & r_y r_z \\ r_z r_x & r_z r_y & r_z^2 \end{bmatrix}$$ (3.40)

As $\widetilde{A} = \widetilde{SC} + \widetilde{SS}$, The strain shear tensor is

$$\widetilde{SS} = \widetilde{A} - \widetilde{SC} = \begin{bmatrix} \frac{\partial u}{\partial x} - \lambda_{cr} & \frac{1}{2}\left(\frac{\partial u}{\partial y} + \frac{\partial v}{\partial x}\right) & \frac{1}{2}\left(\frac{\partial u}{\partial z} + \frac{\partial w}{\partial x}\right) \\ \frac{1}{2}\left(\frac{\partial v}{\partial x} + \frac{\partial u}{\partial y}\right) & \frac{\partial v}{\partial y} - \lambda_{cr} & \frac{1}{2}\left(\frac{\partial v}{\partial z} + \frac{\partial w}{\partial y}\right) \\ \frac{1}{2}\left(\frac{\partial w}{\partial x} + \frac{\partial u}{\partial z}\right) & \frac{1}{2}\left(\frac{\partial w}{\partial y} + \frac{\partial v}{\partial z}\right) & \frac{\partial w}{\partial z} - \lambda_{cr} \end{bmatrix} - (\lambda_r - \lambda_{cr}) \begin{bmatrix} r_x^2 & r_x r_y & r_x r_z \\ r_y r_x & r_y^2 & r_y r_z \\ r_z r_x & r_z r_y & r_z^2 \end{bmatrix}$$ (3.41)

### 3.3.8 2-D principal decomposition in the xy coordinates

In a 2D principal coordinate, the principal decomposition is

$$\nabla\vec{V} = P^T \nabla\vec{v} P = \begin{bmatrix} \lambda_{cr} & -1/2R \\ \epsilon + 1/2R & \lambda_{cr} \end{bmatrix} = R + SC + S = \begin{bmatrix} 0 & -1/2R \\ 1/2R & 0 \end{bmatrix} + \begin{bmatrix} \lambda_{cr} & 0 \\ 0 & \lambda_{cr} \end{bmatrix} + \begin{bmatrix} 0 & 0 \\ \epsilon & 0 \end{bmatrix}$$ (3.42)

Let us come back to the original xy coordinates,

$$\nabla \vec{v} = P \nabla \vec{V} P^T = P \begin{bmatrix} 0 & -1/2R \\ 1/2R & 0 \end{bmatrix} P^T + P \begin{bmatrix} \lambda_{cr} & 0 \\ 0 & \lambda_{cr} \end{bmatrix} P^T + P \begin{bmatrix} 0 & 0 \\ \epsilon & 0 \end{bmatrix} P^T = \widetilde{R} + \widetilde{SC} + \widetilde{S} \quad (3.43)$$

Since Liutex is Galilean invariant,

$$\widetilde{R} = \begin{bmatrix} 0 & -1/2R \\ 1/2R & 0 \end{bmatrix} \quad (3.44)$$

$$\widetilde{SC} = \begin{bmatrix} \lambda_{cr} & 0 \\ 0 & \lambda_{cr} \end{bmatrix} \quad (3.45)$$

$$S = \begin{bmatrix} 0 & 0 \\ \epsilon & 0 \end{bmatrix} = \begin{bmatrix} 0 & 1/2\ \epsilon \\ 1/2\ \epsilon & 0 \end{bmatrix} + \begin{bmatrix} 0 & -1/2\ \epsilon \\ 1/2\ \epsilon & 0 \end{bmatrix} = SS + VS \quad (3.46)$$

$$\nabla \times \vec{V} = \vec{R} + \vec{VS} \text{ and } \vec{VS} = \nabla \times \vec{V} - \vec{R}, \quad (3.47)$$

where $\nabla \times \vec{V}$ is vorticity and $\vec{R}$ is Liutex

The vorticity shear is anti-symmetric and $VS = \begin{bmatrix} 0 & -1/2\ \epsilon \\ 1/2\ \epsilon & 0 \end{bmatrix} = \begin{bmatrix} 0 & -1/2(\omega - R) \\ 1/2(\omega - R) & 0 \end{bmatrix} \quad (3.48)$

The principal decomposition in 2-D xy system can be written as

$$\nabla \vec{v} = \begin{bmatrix} \frac{\partial u}{\partial x} & \frac{\partial u}{\partial y} \\ \frac{\partial v}{\partial x} & \frac{\partial v}{\partial y} \end{bmatrix} = \widetilde{R} + \widetilde{SC} + \widetilde{S} = \begin{bmatrix} 0 & -1/2R \\ 1/2R & 0 \end{bmatrix} + \begin{bmatrix} \lambda_{cr} & 0 \\ 0 & \lambda_{cr} \end{bmatrix} + \begin{bmatrix} \frac{\partial u}{\partial x} - \lambda_{cr} & \frac{\partial u}{\partial y} + 1/2R \\ \frac{\partial v}{\partial x} - 1/2R & \frac{\partial v}{\partial y} - \lambda_{cr} \end{bmatrix} \quad (3.49)$$

The strain shear $\widetilde{SS} = \widetilde{S} - \widetilde{VS} = \begin{bmatrix} \frac{\partial u}{\partial x} - \lambda_{cr} & \frac{\partial u}{\partial y} + \frac{1}{2}\omega \\ \frac{\partial v}{\partial x} - \frac{1}{2}\omega & \frac{\partial v}{\partial y} - \lambda_{cr} \end{bmatrix} = \begin{bmatrix} \frac{\partial u}{\partial x} - \lambda_{cr} & \frac{1}{2}(\frac{\partial u}{\partial y} + \frac{\partial v}{\partial x}) \\ \frac{1}{2}(\frac{\partial u}{\partial y} + \frac{\partial v}{\partial x}) & \frac{\partial v}{\partial y} - \lambda_{cr} \end{bmatrix} \quad (3.50)$

Note that $\omega = \frac{\partial v}{\partial x} - \frac{\partial u}{\partial y}$.

4. **Mathematical check**

This work establish a principal decomposition of the velocity gradient tensor in the original xyz coordinates,

First, we assume there is rotation.

$$\nabla \vec{v} = Q \nabla \vec{V} Q^T = \widetilde{R} + \widetilde{S} + \widetilde{SC} \quad (4.1)$$

$$\begin{bmatrix} \frac{\partial u}{\partial x} & \frac{\partial u}{\partial y} & \frac{\partial u}{\partial z} \\ \frac{\partial v}{\partial x} & \frac{\partial v}{\partial y} & \frac{\partial v}{\partial z} \\ \frac{\partial w}{\partial x} & \frac{\partial w}{\partial y} & \frac{\partial w}{\partial z} \end{bmatrix} = \frac{1}{2} \begin{bmatrix} 0 & -R_z & R_y \\ R_z & 0 & -R_x \\ -R_y & R_x & 0 \end{bmatrix} + \widetilde{S} + \begin{bmatrix} \lambda_{cr} & 0 & 0 \\ 0 & \lambda_{cr} & 0 \\ 0 & 0 & \lambda_{cr} \end{bmatrix} + (\lambda_r - \lambda_{cr}) \begin{bmatrix} r_x^2 & r_x r_y & r_x r_z \\ r_y r_x & r_y^2 & r_y r_z \\ r_z r_x & r_z r_y & r_z^2 \end{bmatrix} \quad (4.2)$$

Mathematical check should prove $\nabla \vec{V} = Q^T \nabla \vec{v} Q$. $\quad (4.3)$

First, in the xyz coordinate, $Q^T \begin{bmatrix} r_x \\ r_y \\ r_z \end{bmatrix} = \begin{bmatrix} 0 \\ 0 \\ 1 \end{bmatrix}$.

Following orthogonal (rotation) matrix Q can satisfy $Q^T \begin{bmatrix} r_x \\ r_y \\ r_z \end{bmatrix} = \begin{bmatrix} 0 \\ 0 \\ 1 \end{bmatrix}$. (4.5)

Proof. Since Q is the orthogonal matrix, $Q_i \cdot Q_j = \begin{cases} 0 & if\ i \neq j \\ 1 & if\ i = j \end{cases}$

Let $Q^T = \begin{bmatrix} Q_{11} & Q_{12} & Q_{13} \\ Q_{21} & Q_{22} & Q_{23} \\ r_x & r_y & r_z \end{bmatrix}$, $Q = \begin{bmatrix} Q_{11} & Q_{21} & r_x \\ Q_{12} & Q_{22} & r_y \\ Q_{13} & Q_{32} & r_z \end{bmatrix}$, (4.6)

$Q^T \begin{bmatrix} r_x \\ r_y \\ r_z \end{bmatrix} = \begin{bmatrix} 0 \\ 0 \\ 1 \end{bmatrix}$.

Since Liutex tensor is anti-symmetric, and Liutex vector is Galilean invariant and parallel to Z in the principal coordinates, the Liutex tensor must be

$$R = \begin{bmatrix} 0 & -1/2R & 0 \\ 1/2R & 0 & 0 \\ 0 & 0 & 0 \end{bmatrix}$$ (4.7)

**Theorem 3.** $Q^T (\lambda_r - \lambda_{cr}) \begin{bmatrix} r_x^2 & r_x r_y & r_x r_z \\ r_y r_x & r_y^2 & r_y r_z \\ r_z r_x & r_z r_y & r_z^2 \end{bmatrix} Q = \begin{bmatrix} 0 & 0 & 0 \\ 0 & 0 & 0 \\ 0 & 0 & \lambda_r - \lambda_{cr} \end{bmatrix}$ if $\begin{bmatrix} r_x \\ r_y \\ r_z \end{bmatrix}$ is the eigenvector of $\nabla \vec{v}$

Proof: $Q^T \begin{bmatrix} r_x^2 & r_x r_y & r_x r_z \\ r_y r_x & r_y^2 & r_y r_z \\ r_z r_x & r_z r_y & r_z^2 \end{bmatrix} Q = \begin{bmatrix} Q_{11} & Q_{12} & Q_{13} \\ Q_{21} & Q_{22} & Q_{23} \\ r_x & r_y & r_z \end{bmatrix} \begin{bmatrix} r_x^2 & r_x r_y & r_x r_z \\ r_y r_x & r_y^2 & r_y r_z \\ r_z r_x & r_z r_y & r_z^2 \end{bmatrix} \begin{bmatrix} Q_{11} & Q_{21} & r_x \\ Q_{12} & Q_{22} & r_y \\ Q_{13} & Q_{32} & r_z \end{bmatrix}$

$= \begin{bmatrix} Q_{11} & Q_{12} & Q_{13} \\ Q_{21} & Q_{22} & Q_{23} \\ r_x & r_y & r_z \end{bmatrix} \begin{bmatrix} 0 & 0 & r_x \\ 0 & 0 & r_y \\ 0 & 0 & r_z \end{bmatrix} = \begin{bmatrix} 0 & 0 & 0 \\ 0 & 0 & 0 \\ 0 & 0 & 1 \end{bmatrix}$ (4.8)

$SC = Q^T \widetilde{SC} Q = Q^T \begin{bmatrix} \lambda_{cr} & 0 & 0 \\ 0 & \lambda_{cr} & 0 \\ 0 & 0 & \lambda_{cr} \end{bmatrix} Q + (\lambda_r - \lambda_{cr}) Q^T \begin{bmatrix} r_x^2 & r_x r_y & r_x r_z \\ r_y r_x & r_y^2 & r_y r_z \\ r_z r_x & r_z r_y & r_z^2 \end{bmatrix} Q = \begin{bmatrix} \lambda_{cr} & 0 & 0 \\ 0 & \lambda_{cr} & 0 \\ 0 & 0 & \lambda_r \end{bmatrix}$ (4.9)

Since $\nabla \vec{v} = Q \nabla \widetilde{V} Q^T = \widetilde{R} + \widetilde{S} + \widetilde{SC}$, $Q^T \widetilde{S} Q = Q^T (\nabla \vec{v} - \widetilde{R} - \widetilde{SC}) Q = \nabla \vec{V} - R - SC = S$

Next, we assume there does not exist rotation.

From Eq. 3.25

$$\widetilde{SC} = Q \begin{bmatrix} \lambda_1 & 0 & 0 \\ 0 & 0 & 0 \\ 0 & 0 & 0 \end{bmatrix} Q^T + Q \begin{bmatrix} 0 & 0 & 0 \\ 0 & \lambda_2 & 0 \\ 0 & 0 & 0 \end{bmatrix} Q^T + Q \begin{bmatrix} 0 & 0 & 0 \\ 0 & 0 & 0 \\ 0 & 0 & \lambda_3 \end{bmatrix} Q^T$$

So,

$$\begin{aligned} Q^T \widetilde{SC} Q &= Q^T \left( Q \begin{bmatrix} \lambda_1 & 0 & 0 \\ 0 & 0 & 0 \\ 0 & 0 & 0 \end{bmatrix} Q^T + Q \begin{bmatrix} 0 & 0 & 0 \\ 0 & \lambda_2 & 0 \\ 0 & 0 & 0 \end{bmatrix} Q^T + Q \begin{bmatrix} 0 & 0 & 0 \\ 0 & 0 & 0 \\ 0 & 0 & \lambda_3 \end{bmatrix} Q^T \right) Q \\ &= \begin{bmatrix} \lambda_1 & 0 & 0 \\ 0 & 0 & 0 \\ 0 & 0 & 0 \end{bmatrix} + \begin{bmatrix} 0 & 0 & 0 \\ 0 & \lambda_2 & 0 \\ 0 & 0 & 0 \end{bmatrix} + \begin{bmatrix} 0 & 0 & 0 \\ 0 & 0 & 0 \\ 0 & 0 & \lambda_3 \end{bmatrix} \\ &= \begin{bmatrix} \lambda_1 & 0 & 0 \\ 0 & \lambda_2 & 0 \\ 0 & 0 & \lambda_3 \end{bmatrix} = SC \end{aligned} \quad (4.10)$$

Similarly, $\nabla \vec{v} = Q \nabla \vec{V} Q^T = \widetilde{R} + \widetilde{S} + \widetilde{SC}$, $Q^T \widetilde{S} Q = Q^T (\nabla \vec{v} - \widetilde{R} - \widetilde{SC}) Q = \nabla \vec{V} - R - SC = S$

The principal decomposition in the original xyz decomposition is proved correct which is same as the principal decomposition in the principal coordinate.

5. Examples in boundary layer transition

In this section, the data of a Direct Numerical Simulation (DNS) is used to verify that the given velocity gradient tensor decomposition is correct. This DNS research is done by Liu et al. in 2014, using the grid level $1920 \times 128 \times 241$ which represent the total amount in streamwise(x), spanwise(y) and wall-normal(z) directions. The detailed information of this DNS can be found in [44].

A point shown in Figure 4, is picked to do the analysis.

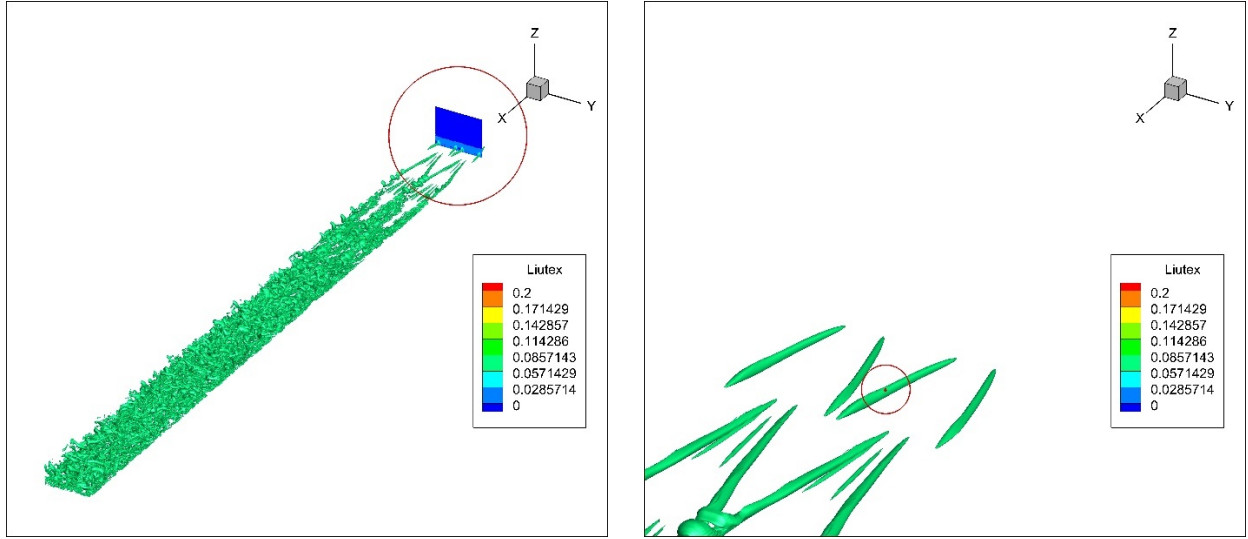

Fig 4 Picked point in the leg region with Liutex isosurface R=0.07

The velocity gradient tensor $\nabla \vec{v}$ in the original coordinate xyz is

$$\nabla \vec{v} = \begin{bmatrix} 0.035937364824172 & 0.125594710923238 & 0.087392848914314 \\ 0.011537008453359 & 0.001192696075516 & 0.205254490413245 \\ 0.009880980182694 & 0.046606547096218 & 0.087392848914314 \end{bmatrix} \tag{5.1}$$

Its eigenvalues are

$$\lambda_1 = 0.020523352560700 \tag{5.2}$$

$$\lambda_2 = 0.051999778626651 + 0.101627185946942i \tag{5.3}$$

$$\lambda_3 = 0.051999778626651 - 0.101627185946942i \tag{5.4}$$

So

$$\lambda_r = 0.020523352560700 \tag{5.5}$$

$$\lambda_{cr} = 0.051999778626651 \tag{5.6}$$

where $\lambda_r$ is the real eigenvalue and $\lambda_{cr}$ is the real part of the complex conjugate eigenvalues.

Easy to know, the vorticity $\vec{\omega}$ is

$$\vec{\omega} = (-0.251861037509463, 0.097273829097008, -0.137131719376597) \tag{5.7}$$

And Liutex $\vec{R}$ is

$$\vec{R} = (-0.0916303471, 0.0139173856, -0.0038396616) \tag{5.8}$$

The classical C-S decomposition will be

$$\nabla \vec{v} = A + B$$
$$= \begin{bmatrix} 0.035937364824172 & 0.057028851234940 & 0.038755934365810 \\ 0.057028851234940 & 0.001192696075516 & 0.079323971658513 \\ 0.038755934365810 & 0.079323971658513 & 0.087392848914314 \end{bmatrix}$$
$$+ \begin{bmatrix} 0 & 0.068565859688299 & 0.048636914548504 \\ -0.068565859688299 & 0 & 0.125930518754732 \\ -0.048636914548504 & -0.125930518754732 & 0 \end{bmatrix} \tag{5.9}$$

Then, compute the velocity gradient tensor in the Principal Coordinate by doing $Q$ and $P$ rotation.(see appendix). The velocity gradient tensor $\nabla \vec{V}$ in the Principal Coordinate will be

$$\nabla \vec{V} = \begin{bmatrix} 0.051999777297995 & 0.046380367062595 & -1.231166141068296 \times 10^{-9} \\ 0.222682253759753 & 0.051999777297995 & -9.732397574264340 \times 10^{-10} \\ 0.033160270185412 & 0.134933282457895 & 0.020523355218012 \end{bmatrix} \tag{5.10}$$

$(\nabla \vec{V})_{13}$ and $(\nabla \vec{V})_{23}$ are supposed to be 0 theoretically, but they are nonzero here due to the numerical error. Since their values are too small, we assume they are 0.

$$\nabla \vec{V} = \begin{bmatrix} 0.051999777297995 & -0.046380367062595 & 0 \\ 0.222682253759753 & 0.051999777297995 & 0 \\ 0.033160270185412 & 0.134933282457895 & 0.020523355218012 \end{bmatrix} \tag{5.11}$$

Easy to know, the Principal Decomposition in Principal Coordinate is

$$\nabla \vec{V} = C + SC + S \tag{5.12}$$

where

$$R = \begin{bmatrix} 0 & -0.046380367062595 & 0 \\ 0.046380367062595 & 0 & 0 \\ 0 & 0 & 0 \end{bmatrix} \tag{5.13}$$

$$SC = \begin{bmatrix} 0.051999777297995 & 0 & 0 \\ 0 & 0.051999777297995 & 0 \\ 0 & 0 & 0.020523355218012 \end{bmatrix} \tag{5.14}$$

$$S = \begin{bmatrix} 0 & 0 & 0 \\ 0.176301886697158 & 0 & 0 \\ 0.033160270185412 & 0.134933282457895 & 0 \end{bmatrix} \tag{5.15}$$

According to the Eq. (3.12), (3.16) and (3.19), the Principal Decomposition in the original coordinate is

$$\widetilde{R} = \frac{1}{2} \begin{bmatrix} 0 & -R_z & R_y \\ R_z & 0 & -R_x \\ -R_y & R_x & 0 \end{bmatrix}$$
$$= \begin{bmatrix} 0 & 0.001919830800000 & 0.006958692800000 \\ -0.001919830800000 & 0 & 0.045815173550000 \\ -0.006958692800000 & -0.045815173550000 & 0 \end{bmatrix} \tag{5.16}$$

$$\widetilde{SC} = \begin{bmatrix} \lambda_{cr} & 0 & 0 \\ 0 & \lambda_{cr} & 0 \\ 0 & 0 & \lambda_{cr} \end{bmatrix} + (\lambda_r - \lambda_{cr}) \begin{bmatrix} r_x^2 & r_x r_y & r_x r_z \\ r_y r_x & r_y^2 & r_y r_z \\ r_z r_x & r_z r_y & r_z^2 \end{bmatrix}$$

$$= \begin{bmatrix} 0.021285836888073 & 0.004665024023157 & -0.001287031495685 \\ 0.004665024023157 & 0.051291225835808 & 1.954823284173003 \times 10^{-4} \\ -0.001287031495685 & 1.954823284173003 \times 10^{-4} & 0.051945847090120 \end{bmatrix} \quad (5.17)$$

$$\widetilde{S} = \begin{bmatrix} 0.014651527936098 & 0.119009856100081 & 0.081721187609999 \\ -0.014282201676516 & -0.050098529760292 & 0.159243834534827 \\ -0.001635255887009 & -9.868558746356920 \times 10^{-4} & 0.035447001824194 \end{bmatrix} \quad (5.18)$$

To verify whether these formulas are correct, $\widetilde{R}$, $\widetilde{SC}$ and $\widetilde{S}$ are transferred into principal coordinate by doing the same $Q$ and $P$ rotation.

$P^T Q^T \widetilde{R} P Q =$
$$\begin{bmatrix} 0 & -0.046380376057143 & 1.374292836139727 \times 10^{-17} \\ 0.046380376057143 & 3.469446951953614 \times 10^{-18} & 6.653842036321013 \times 10^{-18} \\ -1.460093257614954 \times 10^{-17} & -6.767611962670890 \times 10^{-18} & -7.094962483051064 \times 10^{-21} \end{bmatrix}$$

(5.19)

$P^T Q^T \widetilde{SC} P Q$
$$= \begin{bmatrix} 0.051999778626651 & -1.734723475976807 \times 10^{-18} & 4.546300047604554 \times 10^{-18} \\ -1.734723475976807 \times 10^{-18} & 0.051999778626651 & -9.728669630211971 \times 10^{-18} \\ 4.757197723905711 \times 10^{-18} & -9.677239844036469 \times 10^{-18} & 0.020523352560700 \end{bmatrix}$$

(5.20)

$P^T Q^T \widetilde{S} P Q$
$$= \begin{bmatrix} -1.328655869328482 \times 10^{-9} & 8.994547889555647 \times 10^{-9} & -1.231166162874690 \times 10^{-9} \\ 0.176301877702610 & -1.328655831189796 \times 10^{-9} & -9.732397554889825 \times 10^{-10} \\ 0.033160270185412 & 0.134933282457895 & 2.657311829346726 \times 10^{-9} \end{bmatrix}$$

(5.21)

where $P$ and $Q$ are the rotation matrices (see appendix).

Compare Eq. (5.19)-(5.21) with (5.13)-(5.15), it is found that $P^T Q^T \widetilde{R} P Q = R$, $P^T Q^T \widetilde{SC} P Q = SC$ and $P^T Q^T \widetilde{S} P Q = S$, if the numerical errors are ignored which verifies the correctness of Eq. (5.19)-(5.21).

To provide more evidence, the analysis of a point shown in Figure 5 in the ring region is presented in the following.

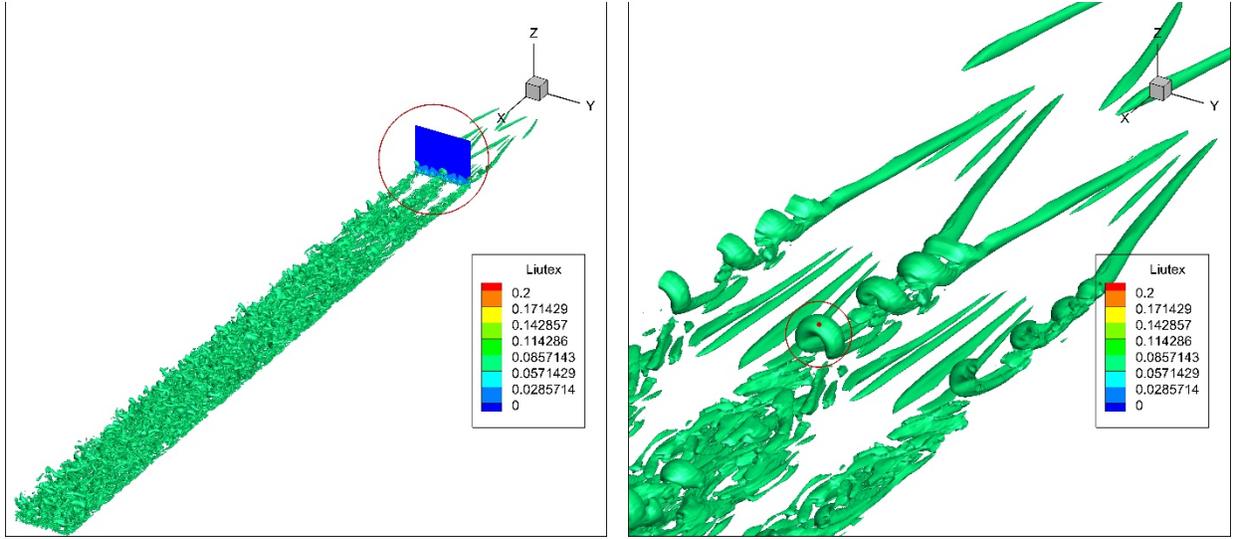

Fig 5 Picked point in the ring region with Liutex isosurface R=0.07

The velocity gradient tensor $\nabla \vec{v}$ in the original coordinate xyz is

$$\nabla \vec{v} = \begin{bmatrix} 0.091017789243287 & -0.009241138132630 & 0.428821405397308 \\ 0.001435173757336 & -0.037696248000415 & -0.002368346237888 \\ -0.141078929092170 & -0.001167284355951 & 0.428821405397308 \end{bmatrix} \quad (5.22)$$

Its eigenvalues are

$$\lambda_1 = 0.259911196546074 + 0.178816503107701i \quad (5.23)$$

$$\lambda_2 = 0.259911196546074 - 0.178816503107701i \quad (5.24)$$

$$\lambda_3 = -0.037679446451969 \quad (5.25)$$

So

$$\lambda_r = -0.037679446451969 \quad (5.26)$$

$$\lambda_{cr} = 0.259911196546074 \quad (5.27)$$

where $\lambda_r$ is the real eigenvalue and $\lambda_{cr}$ is the real part of the complex conjugate eigenvalues.

Like what is done for the first point, vorticity and Liutex can be expressed by

$$\vec{\omega} = (0.001201061881937, 0.569900334489478, 0.010676311889966) \quad (5.28)$$

$$\vec{R} = (0.00398814213, 0.126156181, 0.00152176141) \quad (5.29)$$

The classical C-S decomposition is

$$\nabla \vec{v} = A + B$$

$$= \begin{bmatrix} 0.091017789243287 & -0.003902982187647 & 0.143871238152569 \\ -0.003902982187647 & -0.037696248000415 & -0.001767815296920 \\ 0.143871238152569 & -0.001767815296920 & 0.428821405397308 \end{bmatrix}$$

$$+ \begin{bmatrix} 0 & -0.005338155944983 & 0.284950167244739 \\ 0.005338155944983 & 0 & -6.005309409683390 \times 10^{-4} \\ -0.284950167244739 & 6.005309409683390e-04 & 0 \end{bmatrix} \quad (5.30)$$

The velocity gradient tensor $\nabla \vec{V}$ in the Principal Coordinate is

$$\nabla \vec{V} = \begin{bmatrix} 0.259911196540077 & -0.063114186613453 & -1.342813631698469 \times 10^{-10} \\ 0.506626853674249 & 0.259911196540077 & -4.047833368778066 \times 10^{-10} \\ -0.013663096321051 & -0.010501631274044 & -0.037679446439975 \end{bmatrix} \quad (5.31)$$

And the Principal Decomposition in principal coordinate is

$$\nabla \vec{V} = C + SC + S \quad (5.32)$$

where

$$R = \begin{bmatrix} 0 & -0.063114186613453 & 0 \\ 0.063114186613453 & 0 & 0 \\ 0 & 0 & 0 \end{bmatrix} \quad (5.33)$$

$$SC = \begin{bmatrix} 0.259911196540077 & 0 & 0 \\ 0 & 0.259911196540077 & 0 \\ 0 & 0 & -0.037679446439975 \end{bmatrix} \quad (5.34)$$

$$S = \begin{bmatrix} 0 & 0 & 0 \\ 0.443512667060796 & 0 & 0 \\ -0.013663096321051 & -0.010501631274044 & 0 \end{bmatrix} \quad (5.35)$$

According to the Eq. (3.12), (3.16) and (3.19), the Principal Decomposition in the original coordinate is

$$\widetilde{R} = \frac{1}{2} \begin{bmatrix} 0 & -R_z & R_y \\ R_z & 0 & -R_x \\ -R_y & R_x & 0 \end{bmatrix}$$

$$= \begin{bmatrix} 0 & -7.608807050000000 \times 10^{-4} & 0.063078090500000 \\ 7.608807050000000 \times 10^{-4} & 0 & -0.001994071065000 \\ -0.063078090500000 & 0.001994071065000 & 0 \end{bmatrix} \quad (5.36)$$

$$\widetilde{SC} = \begin{bmatrix} \lambda_{cr} & 0 & 0 \\ 0 & \lambda_{cr} & 0 \\ 0 & 0 & \lambda_{cr} \end{bmatrix} + (\lambda_r - \lambda_{cr}) \begin{bmatrix} r_x^2 & r_x r_y & r_x r_z \\ r_y r_x & r_y^2 & r_y r_z \\ r_z r_x & r_z r_y & r_z^2 \end{bmatrix}$$

$$= \begin{bmatrix} 0.259614134934160 & -0.009396896414233 & -1.133502474757582 \times 10^{-4} \\ -0.009396896414233 & -0.037339133615098 & -0.003585587943163 \\ -1.133502474757582 \times 10^{-4} & -0.003585587943163 & 0.259867945321118 \end{bmatrix} \quad (5.37)$$

$$\tilde{S} = \begin{bmatrix} -0.168596345690873 & 9.166389866027106 \times 10^{-4} & 0.365856665144784 \\ 0.010071189466569 & -3.571143853177763 \times 10^{-4} & 0.003211312770275 \\ -0.077887488344695 & 4.242325222112934 \times 10^{-4} & 0.168953460076190 \end{bmatrix} \quad (5.38)$$

Transfer $\widetilde{R}$, $\widetilde{SC}$ and $\tilde{S}$ into principal coordinate

$$P^T Q^T \widetilde{R} P Q$$
$$= \begin{bmatrix} 0 & -0.063114188262115 & 5.043701192153824 \times 10^{-18} \\ 0.063114188262115 & 0 & -4.717137691159609 \times 10^{-18} \\ -4.696457779418198 \times 10^{-18} & 4.999408747473241 \times 10^{-18} & -3.306421514649439 \times 10^{-21} \end{bmatrix}$$

(5.39)

$$P^T Q^T \widetilde{SC} P Q$$
$$= \begin{bmatrix} 0.259911196546074 & -2.775557561562891 \times 10^{-17} & -2.319958772685060 \times 10^{-17} \\ -1.387778780781446 \times 10^{-17} & 0.259911196546074 & -2.217472719314102 \times 10^{-17} \\ -2.210933145837840 \times 10^{-17} & -2.453642639070820 \times 10^{-17} & -0.037679446451968 \end{bmatrix}$$

(5.40)

$$P^T Q^T \tilde{S} P Q$$
$$= \begin{bmatrix} -5.996720395714053 \times 10^{-12} & 1.648662175247798 \times 10^{-9} & -1.342813463615500 \times 10^{-10} \\ 0.443512665412133 & -5.996730889634705 \times 10^{-12} & -4.047833081012846 \times 10^{-10} \\ -0.013663096321051 & -0.010501631274044 & 1.199335850901248 \times 10^{-11} \end{bmatrix}$$

(5.41)

where $P$ and $Q$ are the rotation matrices (see appendix).

Like what is found with the first leg point, $P^T Q^T \widetilde{R} P Q = R$, $P^T Q^T \widetilde{SC} P Q = SC$ and $P^T Q^T \tilde{S} P Q = S$, if the numerical errors are ignored.

6. Conclusion

According to our previous work [6][27], the velocity gradient tensor should be decomposed in a Liutex-based principal coordinate to rigid rotation, pure shear and stretching (compression). However, the principal coordinate needs twice coordinate rotations or transformed twice by the rotational matrix, namely $Q$ and $P$ rotations, which is not convenient to carry out. In fact, we only use them as a concept but, in general, never do real $Q$ and $P$ rotations. On the other hand, since each point has its own principal coordinates which is different from each other for every point. We then must use the original xyz- coordinate system to do the tensor principal decomposition. The principal decomposition in a principal coordinate is not practical for the velocity tensor decomposition. This paper gives a detailed derivation and steps to carry out the principal decomposition of velocity gradient tensor in the original xyz system for every point, which will provide a very useful tool for development of new fluid kinematics and further development of the new governing equations for fluid dynamics. Following conclusions can be made.

1. In the Cauchy-Stokes decomposition of velocity gradient, the vorticity tensor cannot represent fluid rotation or vortex. The strain tensor cannot distinguish between stretching (compression) and shear, which is not Galilean invariant.
2. Anti-symmetric tensor or vorticity tensor cannot represent fluid rotation and must be decomposed to a rigid rotation (Liutex) and an anti-symmetric (vorticity) shear
3. A principal coordinate is based on Liutex as the Z-axis and that X- and Y- axes make the two diagonal elements of the velocity gradient tensor equal.
4. The principal decomposition of the velocity tensor can be made in the principal coordinate, which has clear physical meanings as rigid rotation, pure shear and stretching (compression). Moreover, all elements are Galilean invariant and the decomposition is unique and Galilean invariant.
5. Since each point has its own Liutex and therefore its own principal coordinate, the principal decomposition of the velocity gradient tensor must be transformed back to the original xyz-system.
6. The principal decomposition in the original xyz-system has been derived which should be used to replace the traditional Cauchy-Stokes decomposition as CS is not Galilean invariant and the vorticity tensor cannot be used to represent fluid rotation or vortex.

7. Appendix:

Algorithm to find rotation matrices P and Q.

Step1: Rotate around z axis

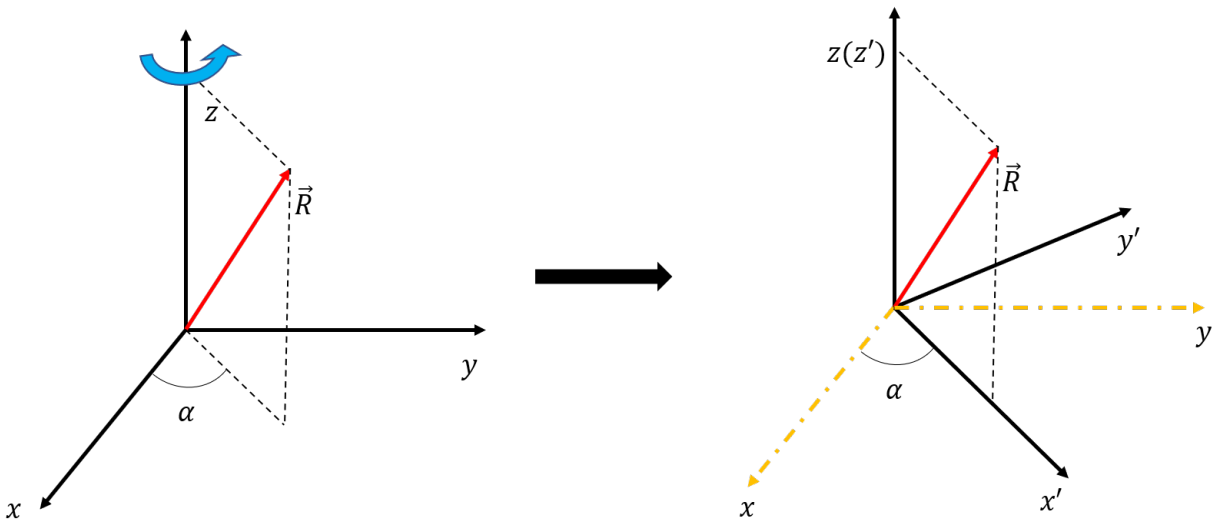

Fig 6 rotate around z-axis

Suppose $\vec{R} = (R_x, R_y, R_z)$. Angle $\alpha$ can be calculated by

$$\alpha = \text{acos}\left(\frac{(R_x, R_y) \cdot (1,0)}{\sqrt{R_x^2 + R_y^2}}\right) \tag{7.1}$$

where $(R_x, R_y)$ is the projection of $\vec{R}$ in xy plane and $(1,0)$ is the direction of x-aixs (Fig 6).

The rotation matrix around z-axis $Q_x$ is

$$Q_x = \begin{bmatrix} \cos\alpha & \sin\alpha & 0 \\ -\sin\alpha & \cos\alpha & 0 \\ 0 & 0 & 1 \end{bmatrix} \tag{7.2}$$

Step 2: Rotate around y' axis.

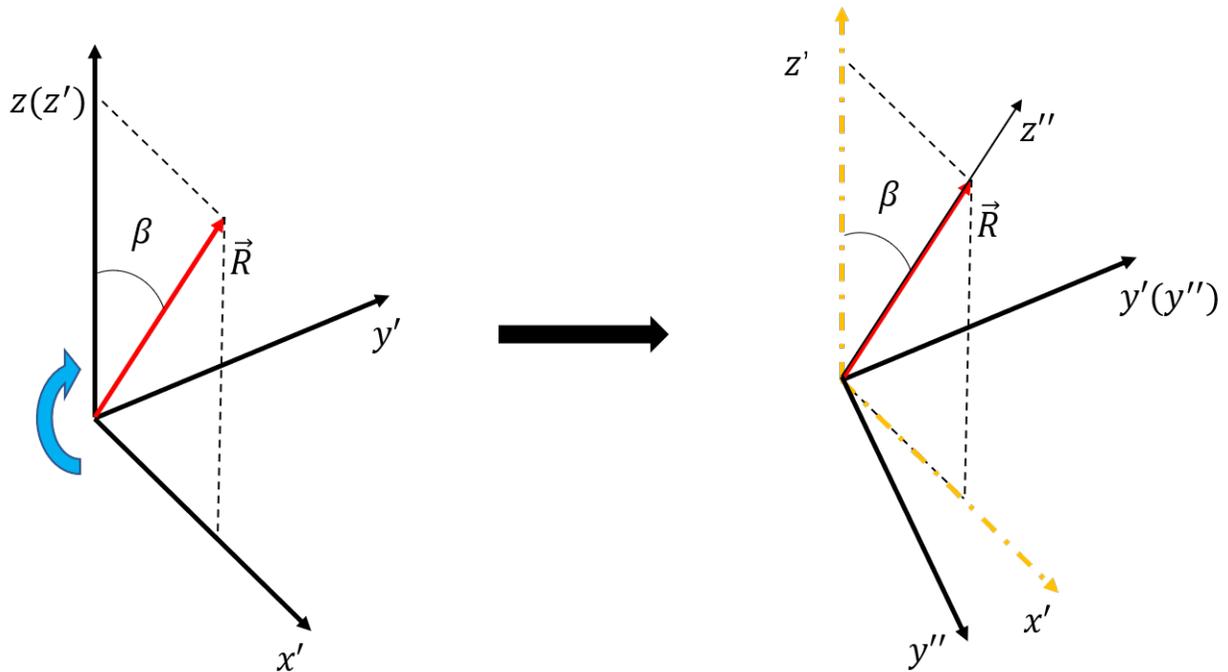

Fig 7 rotate around y'-axis

Angle $\beta$ can be calculated by

$$\beta = \text{acos}\left(\frac{\vec{R} \cdot (0,0,1)}{\sqrt{R_x^2 + R_y^2 + R_z^2}}\right) \tag{7.3}$$

where $(0,0,1)$ is the direction of z-axis. It is noted that the coordinates of $\vec{R}$ and z-axis are both expressed in the original xyz coordinate, since step 1 does not change $\beta$ (Fig 7).

The rotation matrix around y'-axis $Q_{y'}$ is

$$Q_{y'} = \begin{bmatrix} \cos\alpha & 0 & -\sin\alpha \\ 0 & 1 & 0 \\ \sin\alpha & 0 & \cos\alpha \end{bmatrix} \tag{7.4}$$

In general, the rotation matrix Q is

$$Q = Q_{y'}Q_x \tag{7.5}$$

By doing Q rotation, the z''-axis is along the same direction of Liutex $\vec{R}$. The velocity gradient tensor in the x''y''z'' coordinate is

$$\overrightarrow{\nabla v''} = Q(\nabla \vec{v})Q^T \tag{7.6}$$

Step 3: Rotate around y''-axis.

After making z''-axis along the same direction of Liutex $\vec{R}$, the coordinate will be rotated around y''-axis to make the velocity gradient tensor satisfy the part 2 and 3 of the Principal Coordinate definition.

The velocity gradient tensor $\overrightarrow{\nabla v''}$ is in the form of

$$\overrightarrow{\nabla v''} = \begin{bmatrix} (\overrightarrow{\nabla v''})_{11} & (\overrightarrow{\nabla v''})_{12} & 0 \\ (\overrightarrow{\nabla v''})_{21} & (\overrightarrow{\nabla v''})_{22} & 0 \\ (\overrightarrow{\nabla v''})_{31} & (\overrightarrow{\nabla v''})_{32} & \lambda_r \end{bmatrix} \tag{7.7}$$

Suppose it is required to rotate angle $\theta$. The rotation matrix is

$$P = \begin{bmatrix} \cos\theta & \sin\theta & 0 \\ -\sin\theta & \cos\theta & 0 \\ 0 & 0 & 1 \end{bmatrix} \tag{7.8}$$

The velocity gradient tensor after rotation $\nabla \vec{V}$ [27] is

$$\nabla \vec{V} = P\overrightarrow{\nabla v''}P^T = \begin{bmatrix} \frac{\partial U}{\partial X} & \frac{\partial U}{\partial Y} & 0 \\ \frac{\partial V}{\partial X} & \frac{\partial V}{\partial Y} & 0 \\ \frac{\partial W}{\partial X} & \frac{\partial W}{\partial Y} & \lambda_r \end{bmatrix} \tag{7.9}$$

where

$$\frac{\partial U}{\partial Y} = a\sin(2\theta + \varphi) - b \tag{7.10}$$

$$\frac{\partial V}{\partial X} = a\sin(2\theta + \varphi) + b \tag{7.11}$$

$$a = \frac{1}{2}\sqrt{\left((\overrightarrow{\nabla v''})_{22} - (\overrightarrow{\nabla v''})_{11}\right)^2 + \left((\overrightarrow{\nabla v''})_{21} + (\overrightarrow{\nabla v''})_{12}\right)^2} \tag{7.12}$$

$$b = \frac{1}{2}\left((\nabla\overrightarrow{v''})_{21} - (\nabla\overrightarrow{v''})_{12}\right) \tag{7.13}$$

$$\varphi = \begin{cases} acos\left(\dfrac{\frac{1}{2}\left((\nabla\overrightarrow{v''})_{22} - (\nabla\overrightarrow{v''})_{11}\right)}{a}\right), & (\nabla\overrightarrow{v''})_{21} + (\nabla\overrightarrow{v''})_{12} \geq 0 \\ asin\left(\dfrac{\frac{1}{2}\left((\nabla\overrightarrow{v''})_{21} + (\nabla\overrightarrow{v''})_{12}\right)}{a}\right), & (\nabla\overrightarrow{v''})_{21} + (\nabla\overrightarrow{v''})_{12} < 0, (\nabla\overrightarrow{v''})_{22} - (\nabla\overrightarrow{v''})_{11} \geq 0 \\ asin\left(\dfrac{-\frac{1}{2}\left((\nabla\overrightarrow{v''})_{21} + (\nabla\overrightarrow{v''})_{12}\right)}{a}\right) + \pi, & (\nabla\overrightarrow{v''})_{21} + (\nabla\overrightarrow{v''})_{12} < 0, (\nabla\overrightarrow{v''})_{22} - (\nabla\overrightarrow{v''})_{11} < 0 \end{cases} \tag{7.14}$$

Recall that the fundamental Liutex magnitude expression [45] is 2(b-a), so when $2\theta + \varphi = \frac{\pi}{2}$, the velocity gradient tensor satisfies definition of Principal Coordinate. Therefore,

$$\theta = \frac{1}{2}\left(\frac{\pi}{2} - \varphi\right) \tag{7.15}$$

Then the value of each entry is known.

$$P = \begin{bmatrix} cos\theta & sin\theta & 0 \\ -sin\theta & cos\theta & 0 \\ 0 & 0 & 1 \end{bmatrix} \tag{7.16}$$

After that, the velocity gradient tensor in Principal Coordinate is found.

$$\nabla\vec{V} = P\nabla\overrightarrow{v''}P^T = PQ(\nabla\vec{v})Q^T P^T \tag{7.17}$$

**Data availability**

The data that supports the findings of this study are available from the corresponding author upon reasonable request.

**Acknowledgement**

The authors are thankful for the support by the UTA Department of Mathematics which houses the UTA Vortex and Turbulence Research Team. The authors are also grateful to Texas Advanced Computing Center (TACC) for providing computation time. The DNSUTA code was released by Chaoqun Liu in 2009 and the Liutex code was released by Chaoqun Liu in 2018 which can be downloaded from the UTA web site at https://www.uta.edu/math/cnsm/public_html/cnsm/cnsm.html.